\documentclass[12pt]{iopart}
\usepackage{orcidlink}
\usepackage{hyperref}
\usepackage{float}
\usepackage{adjustbox}
\usepackage{hhline}
\usepackage{multirow}
\usepackage{ragged2e}
\usepackage{epsfig}
\usepackage{caption}
\usepackage{latexsym}
\usepackage{bm}
\usepackage{subcaption}
\usepackage{amsfonts}
\usepackage{iopams}
\usepackage{graphicx}
\usepackage{palatino}
\usepackage{mathpazo}
\usepackage{hhline}
\usepackage{multirow}
\usepackage{textcomp}
\usepackage{booktabs}
\usepackage{hhline}
\usepackage{xcolor}
\hypersetup{colorlinks,citecolor=blue}
\hypersetup{colorlinks=true,linkcolor=red,filecolor=magenta, urlcolor=blue}

\begin{document}

\title[Preparing for IOP Publishing journals]{Dynamical system analysis in modified Galileon cosmology}

\author{L.K. Duchaniya$^{1,a}$\orcidlink{0000-0001-6457-2225}, B. Mishra$^{2,a}\footnote{Author to whom any correspondence should be addressed}$ \orcidlink{0000-0001-5527-3565}, I. V. Fomin$^{3,b}$\orcidlink{0000-0003-1527-914X}  and S. V. Chervon$^{4,b,c}$\orcidlink{0000-0001-8898-3694}}

\address{$^a$ Department of Mathematics, Birla Institute of Technology and Science-Pilani, Hyderabad Campus, Hyderabad-500078, India.\\ $^b$ Bauman Moscow State Technical University, Russia.\\ $^c$ Ulyanovsk State Pedagogical University, Ulyanovsk, Russia.}
\ead{$^1$duchaniya98@gmail.com,$^2$bivu@hyderabad.bits-pilani.ac.in, $^3$ingvor@inbox.ru and $^4$chervon.sergey@gmail.com}
\vspace{10pt}

\begin{abstract}
In this paper, we have investigated the phase space analysis in modified Galileon cosmology, where the Galileon term is considered a coupled scalar field, $F(\phi)$. We focus on the exponential type function of $F(\phi)$ and the three well-motivated potential functions $V(\phi)$. We obtain the critical points of the autonomous system, along with their stability conditions and cosmological properties. The critical points of the autonomous system describe different phases of the Universe. The scaling solution for critical points was found in our analysis to determine the matter-dark energy and radiation-dark energy-dominated eras of the Universe. In these scaling solutions, dark energy is typically introduced alongside another component, such as radiation or matter, and helps explain the transition between cosmological eras. The dark energy dominated critical points show stable behavior and indicate the late-time cosmic acceleration phase of the Universe. Further, the results are examined with the Hubble rate $H(z)$ and the Supernovae Ia cosmological data sets.
\end{abstract}

\section{Introduction} \label{SEC I}

In recent years, several cosmological models have been proposed in modified gravity theories to address the issue of late-time cosmic phenomena \cite{SupernovaSearchTeam:1998fmf, SupernovaCosmologyProject:1998vns} and in particular the resolution to Hubble tension issue \cite{Abdalla:2022yfr, DiValentino:2021izs, Brout:2022vxf, Bernal:2016gxb, Benisty:2021cmq}. Though several cosmological problems have been resolved in General Relativity (GR), some key issues still need to be addressed. Therefore, modification/extension in GR is inevitable. Someone can either modify the matter part or the underlying geometry. As part of the first approach, additional fields are introduced into the Universe, including canonical and phantom fields, vector and chiral fields, k-essence, etc. \cite{Copeland:2006wr, Bassett:2005xm, Cai:2009zp, Bahamonde:2017ize, Otalora:2013dsa, Chervon_2019prd, Fomin_2022jcap}. In the second approach, the geometry part is modified to produce modified/extended gravitational theories \cite{Capozziello:2011et}. Modified gravitational theories are typically created by extending the Einstein-Hilbert action, which is curvature-based. However, an alternate approach involves extending the action of the equivalent torsional formulation of GR, the teleparallel equivalent of GR (TEGR) \cite{Maluf:1994ji, Aldrovandi:2013wha, deAndrade:2000nv, Pereira:2019woq, Bahamonde:2021gfp, 2023Symm15291C}.\\

The formalism of the modified gravity theory that produces second-order equations in four-dimensional space-time can originate from TEGR. In TEGR, the metric tensor is replaced by the dynamical variable tetrad. In GR, the Levi-Civita connection incorporates curvature without torsion, whereas in teleparallelism, the Weitzenb$\ddot{o}$ck connection incorporates torsion without curvature \cite{Weitzenbock1923}. 
In this framework, the dynamic elements are the four linearly independent tetrad fields that form the orthogonal bases for the tangent space at each point in space-time. Furthermore, the torsion tensor is derived from the product of the first derivatives of the tetrad fields. The simplest modification of TEGR is $f(T)$ gravity theories, in which the torsion scalar $T$ has been replaced with $f(T)$ \cite{Linder:2010py, Bengochea:2008gz, Cai:2015emx, Farrugia:2016xcw, Finch:2018gkh, Duchaniya:2022rqu, Briffa_2024, Duchaniya_2024ab}. The further extension on $f(T)$ gravity are: $f(T, B)$ gravity \cite{Escamilla-Rivera:2019ulu, Kadam_2023ab} (coupled with boundary term $B$), $f(T, T_{G})$ gravity \cite{Kofinas:2014daa, Lohakare_2023a} (coupled with Gauss-bonnet term $T_{G}$),  $f(T, \phi)$ gravity \cite{Gonzalez_Espinoza_2020A, Gonzalez-Espinoza:2021mwr, Duchaniya_2023noet, Fomin_2024STF}, and non-minimally coupled scalar-torsion gravity \cite{Sahni_2004, Geng:2011ka, Hohmann:2018rwf, Chervon2019a,  Kadam:2022lgq, Paliathanasis:2021nqa} and so on. The other general family of gravitational modifications are called scalar-tensor theories, constructed by coupling curvature parameters to a single scalar field. The most comprehensive four-dimensional scalar-tensor theory with a single propagating scalar degree of freedom is called Horndeski gravity \cite{Horndeski:1974wa} or a similar generalized version of Galileon theory \cite{Nicolis:2008in, Deffayet:2009wt, Leon_2013, Baker:2017hug}.\\

A more interesting case could arise by introducing non-minimal coupling terms and higher derivative quantum gravity modifications in the action. A modification of this might take the form of a field self-interaction of the Galileon type in the form $G(\phi, X)\Box\phi$ where $X=-\partial_ \mu \phi \partial^ \mu \phi/2$. $G$ is an arbitrary function of $\phi$ and kinetic energy $X$ \cite{Nicolis:2008in}. Based on the Dvali-Gabadadze-Porrati (DGP) model \cite{Dvali_2000ga} decoupling limit and its cosmological implications \cite{Deffayet_2001gab}, the Galileon theory was developed. The Galilean shift symmetry, which maintains the equation of motion in the second order, causes the gradient of the scalar field to change constantly from $\partial_{\mu}\phi \rightarrow \partial_{\mu}\phi+c_{\mu}$. It has been demonstrated that the Galilean symmetry cannot be maintained after the theory is covariantized, but the required features may still be maintained. For instance, the scalar field equation of motion can continue to be of second order, which is crucial since higher derivative theories have additional degrees of freedom. In the literature, the Galileon cosmology has been thoroughly examined within the cosmological framework. The late-time cosmic acceleration of the Universe studied in Ref. \cite{Silva_2009, Germani_2012, Easson_2021}. Refs. \cite{Burrage_2011, Renaux_Petel_2013, Gonzalez_Espinoza_2019,  Choudhury_2024, Chaadaeva_2024STF} have explained inflation phase of the Universe. According to Refs. \cite{Burrage_2010, Appleby_2012, Iorio_2012}, the model parameters are constrained by cosmological observations. \\

The analytical approach in framing cosmological models in modified theories of gravity leads to complicated systems of equations and ambiguous initial conditions. This demonstrates the importance of using dynamical system theory for qualitative analysis. Dynamical system analysis is an effective tool for studying the evolution of the Universe through the critical points of the autonomous system. From the stability analysis of the critical points, one can interpret different evolutionary phases of the Universe. In this paper, we conduct a phase-space and stability analysis of Galileon cosmology, methodically examining various cosmological behaviors. The focus is on stable solutions for late time since this field of study displays intriguing phenomenological characteristics. Dynamical dark energy models, including quintessence, phantom, coupled dark energy, modified gravity theories, k-essence, and tracker solutions, generally yield two-component scaling solutions, such as the dark energy-radiation and dark energy-matter eras. These solutions were originally obtained by Wetterich (1988) \cite{Wetterich_1988} and Ferreira and Joyce(1998) 
\cite{Ferreira_1998}. These scaling solutions typically involve the proportional evolution of dark energy alongside another component, such as radiation or matter, and help explain the transition between different cosmological eras. We also calculate various observables in these asymptotic solutions, such as dark energy density, equation-of-state (EoS), and deceleration parameters. Several cosmological systems in modified theories of gravity may be seen in Refs. \cite{Amendola_2000_cup, Roy_2018pha, Gonzalez-Espinoza:2020jss, Rave_Franco_2021jackdynacom,  Kadam:2022lgq, Narawade_2022a, Duchaniya_2023tphi, Patil_2023,  Lohakare_2023_40, Agrawal_2024, Duchaniya_2024tt, Kadam_2024ttg, Gonzalez_Espinoza_2024pha, PhysRevD.109.124044, Chervon_2023GC, Chervon_2020Uni, Vishwakarma_2024}.
 \\

The paper is organized as follows: in Section--\ref{SECII}, we provide a brief overview of the mathematical formalism of torsion Galileon theory. The dynamical system analysis is in Section--\ref{SECIII}. It has three subsections for three different well-motivated forms of $V(\phi)$. Results and discussion are given in Section--\ref{SECIV}. In Section--\ref{Sec-app}, the description of Central Manifold Theory has been given, whereas in Section--\ref{dataset} the details of cosmological datasets are provided.

\section{Mathematical formalism}\label{SECII} 
In GR, gravity is described by curvature using the Levi-Civita connection \cite{Clifton:2011jh}, $\mathring{\Gamma}^{\sigma}_{\;\mu \nu}$ ( over circle represents quantities determined by the Levi-Civita connection), within the framework of Riemann geometry. In contrast to curvature-based gravity explanations, the idea of TG \cite{Hayashi:1979qx, Aldrovandi:2013wha} has been considered. Here the Weitzenb$\ddot{o}$ck connection $\Gamma^{\sigma}_{\;\mu \nu}$ takes the role of the Levi-Civita connection. Teleparallel theories of gravity can be accomplished by using the tetrad $e^{A}_{\,\,\,\,\mu}$ (and its inverses $E^{\,\,\,\, \mu}_{A}$), which replaces the metric as the fundamental variable of gravity theory.
\begin{eqnarray}\label{metric_tetrad_rel}
    g_{\mu\nu}=e^{A}_{\,\,\,\,\mu} e^{B}_{\,\,\,\,\nu}\eta_{AB}\,,\hspace{2cm}\eta_{AB} = E^{\,\,\,\, \mu}_{A} E^{\,\,\,\, \nu}_{B}g_{\mu\nu}\,.
\end{eqnarray}
The Latin indices refer to coordinates on the general manifold, while the Greek indices refer to coordinates on the tangent space. The orthogonality conditions for tetrads can described as
\begin{eqnarray}
 e^{A}_{\,\,\,\,\mu} E_{B}^{\,\,\,\,\mu}=\delta^A_B\,,\hspace{2cm} e^{A}_{\,\,\,\,\mu} E_{A}^{\,\,\,\,\nu}=\delta^{\nu}_{\mu}\,.    
\end{eqnarray}
The Weitzenb$\ddot{o}$ck connection can be defined as \cite{Weitzenbock1923,Krssak:2015oua} 
\begin{equation}
    \Gamma^{\sigma}_{\,\,\,\,\nu\mu} := E_{A}^{\,\,\,\,\sigma}\left(\partial_{\mu}e^{A}_{\,\,\,\,\nu} + \omega^{A}_{\,\,\,\,B\mu} e^{B}_{\,\,\, \nu}\right)\,.
\end{equation}
For the Weitzenb$\ddot{o}$ck gauge, the spin connection preserves the local Lorentz invariance, and these components vanish identically. This connection allows us to define \cite{Hayashi:1979qx} as an anti-symmetric operator on the Riemann tensor, which vanishes for the teleparallel connection. 
\begin{equation}
    T^{\sigma}_{\,\,\,\,\mu\nu} :=2\Gamma^{\sigma}_{\,\,\,[\nu\mu]}\,,
\end{equation}
this tensor is covariant under local Lorentz transformations and diffeomorphisms, where the square brackets stand for the anti-symmetry operators. For torsion scalars, the form obtained by combining contractions of torsion tensors is as follows\cite{Krssak:2018ywd, Bahamonde:2021gfp}
\begin{equation}\label{torsionscalar}
T \equiv \frac{1}{4} T^{\rho \mu \nu} T_{\rho \mu \nu}+\frac{1}{2} T^{\rho \mu \nu} T_{\nu \mu \rho}-T_{\rho \mu}^{~~\rho} T^{\nu \mu}_{~~\nu}.
\end{equation}
The linear form of $T$ describes the action of TEGR. Thus, we can define the action of TEGR as \cite{Maluf:1994ji}
\begin{equation}\label{TEGR_Lagran} 
    \mathcal{S}_{TEGR}^{} =  -\frac{1}{2\kappa^2}\int \mathrm{d}^4 x\; e T  + \int \mathrm{d}^4 x\; e (\mathcal{L}_{m}+\mathcal{L}_{r})\,.
\end{equation}   
This work will study the phase space analysis in torsion-Galileon theory. We briefly review the mathematical formalism of the Torsion-Galileon cosmology through the teleparallel Horndeski theory of formalism. Horndeski theory of gravity is the most comprehensive explanation of gravity in a four-dimensional spacetime \cite{Horndeski:1974wa}. Lovelock's theorem \cite{Lovelock:1971yv} limits the use of Lagrangian terms to create second-order theories. The Lagrangian represents the Teleparallel analog of Horndeski theory of gravity \cite{Bahamonde:2019shr} as, 
\begin{equation}\label{horn_lag}
\mathcal{L}=\sum_{i=2}^5 \mathcal{L}_i \,,    
\end{equation}
this is the most general theory with a single scalar field comprising scalar invariants, which are at most quadratic in torsion tensor contractions. Further, it leads to second-order field equations regarding derivatives of the tetrad or scalar field. The Lagrangians $\mathcal{L}_{i}$, for $i=2,3,4, 5$ \cite{Horndeski:1974wa} respectively represents
\begin{eqnarray}
 \mathcal{L}_{2} & :=G_{2}(\phi,X)\,,\label{eq:LagrHorn1}\\
    \mathcal{L}_{3} & :=G_{3}(\phi,X)\Box\phi\,,\\
    \mathcal{L}_{4} & :=G_{4}(\phi,X)\left(-T+B\right)+G_{4,X}(\phi,X)\left(\left(\Box\phi\right)^{2}-\phi_{;\mu\nu}\phi^{;\mu\nu}\right)\,,\\
     \mathcal{L}_{5} & :=G_{5}(\phi,X)G_{\mu\nu}\phi^{;\mu\nu}-\frac{1}{6}G_{5,X}(\phi,X)\bigg(\left(\Box\phi\right)^{3}+2 \phi_{;\mu}^{~~\nu} \phi_{;\nu}^{~~\alpha}\phi_{;\alpha}^{~~\mu}\nonumber\\  &-3\phi_{;\mu\nu}\phi^{;\mu\nu}\,\Box\phi\bigg)\,.\label{eq:LagrHorn5}    
\end{eqnarray}
Here, $\phi$ and $X$ denote the scalar field and kinetic energy. Also $G_{i}$, $i=2,3,4,5$ is a function of scalar field $\phi$ and kinetic energy $X=-\partial_ \mu \phi \partial^ \mu \phi/2$. Whereas $G_{i, X}$ and $G_{i,\phi}$ respectively denote the partial derivatives of $G_{i}$ with respect to $X$ and $\phi$. \\
The matter and radiation content of the Universe respectively represented by the Lagrangian $\mathcal{L}_{m}$, $\mathcal{L}_{r}$. This is equivalent to the perfect fluid with an energy density of $\rho_m$, $\rho_r$, and pressure of $p_m$, $p_r$. This exists in addition to the scalar-tensor sectors mentioned earlier. The entire action can be written as 
\begin{eqnarray}\label{Horndeski_action_Lagran}
    \mathcal{S} =  \int \mathrm{d}^4 x e\, (\mathcal{L}+\mathcal{L}_m+\mathcal{L}_r) \,,
\end{eqnarray}
where $e$ is the determinant of the tetrad, $e=Det[e^{A}_{\,\,\,\,\mu}]=\sqrt{-g}$.\\ 
The Galileon emerges in its first form, non-minimally coupled to matter. So, we strictly regard the Galileon theory in the present study not as a modified version of gravity but as a scalar field and dark-energy construction. \\

We consider homogeneous and isotropic flat Friedmann-Lema\^{i}tre-Robertson-Walker (FLRW) space time,
\begin{equation}\label{metric}
ds^{2} = -N(t)^2 dt^{2}+a(t)^2[dx^2+dy^2+dz^2]\,,    
\end{equation}  
where $N$ and $a(t)$ respectively be the lapse function and scale factor and $T=6H^{2}$. Varying the action (\ref{Horndeski_action_Lagran}) with respect to lapse function $N$, the first Friedmann equation can be obtained as \cite{Bahamonde:2019shr},
\begin{eqnarray}\label{first_friedmann_equation}     
&2X G_{2,X}-G_{2}+6X\dot{\phi}HG_{3,X}-2XG_{3,\phi}-6H^{2}G_{4}+24H^{2}X(G_{4,X}+XG_{4,XX}) \nonumber\\&-12HX\dot{\phi}G_{4,\phi X}-6H\dot{\phi}G_{4,\phi} +2H^{3}X \dot{\phi}(5G_{5,X}+2XG_{5,XX})\nonumber\\&-6H^{2}X(3G_{5,\phi}+2XG_{5,\phi X})=\rho_{m}+\rho_{r}\,,    
\end{eqnarray}  
and varying the action (\ref{Horndeski_action_Lagran}) with respect to scale factor $a(t)$, one can obtain
\begin{eqnarray}\label{second_friedmann_equation}
&G_{2}-2X(G_{3, \phi}+\ddot{\phi}G_{3, X})+2G_{4}(3H^{2}+2\dot{H})-12H^{2}XG_{4, X}-4H\dot{X}G_{4, X}\nonumber\\&-8\dot{H}XG_{4,X}-8HX\dot{X}G_{4, XX}+2G_{4, \phi}(\ddot{\phi}+2H\dot{\phi})+4XG_{4, \phi \phi}+4X(\ddot{\phi}\nonumber\\&-2H\dot{\phi})G_{4, \phi X}-2X(2H^{3}\dot{\phi}+2H\dot{H}\dot{\phi}+3H^{2}\ddot{\phi})G_{5, X}-4H^{2}X^{2} \ddot{\phi} G_{5, XX}\nonumber\\&+4HX(\dot{X}-HX)G_{5,\phi X} + 2\left[2\frac{d}{dt}(HX)+3H^{2}X\right]\nonumber\\&+4HX\dot{\phi}G_{5,\phi \phi}=-p_{m}-p_{r}\,.    
\end{eqnarray}
Now taking variation of (\ref{Horndeski_action_Lagran}) with respect to scalar field $\phi$, we can get the Klein-Gordan equation 
\begin{equation}\label{Klein-Gordan-equation}
\frac{1}{a^{3}}\frac{d}{dt}(a^{3}J)=P_{\phi}\,,    
\end{equation}
where
\begin{eqnarray}
J&=& \dot{\phi}G_{2, X}+6HXG_{3, X}-2\dot{\phi}G_{3, \phi}+6H^{2}\dot{\phi}(G_{4, X}+2XG_{4, XX})-12HXG_{4,\phi X}  \nonumber \\ &&+2H^{3}X(3G_{5,X}+2XG_{5, XX})-6H^{2}\dot{\phi}(G_{5,\phi}+XG_{5, \phi X})\,, \nonumber \\
P_{\phi}&=& G_{2, \phi}-2X(G_{3, \phi \phi}+\ddot{\phi}G_{3,\phi X})+6G_{4, \phi}(2H^{2}+\dot{H})+6HG_{4,\phi X}(\dot{X}+2HX)\nonumber \\ &&-6H^{2}XG_{5,\phi \phi}+2H^{3}X \dot{\phi} G_{5, \phi X}\,.
\end{eqnarray}
To study the phase space analysis in the torsion Galileon theory, we need to define the function $G_{i}$, $i = 2, 3, 4, 5$ as
\begin{eqnarray} \label{G_i_functions}
G_{2}(\phi, X)&=&X-V(\phi)\,,\nonumber\\
G_{3}(\phi, X)&=&-F(\phi)X\,,\nonumber\\
G_{4}(\phi, X)&=&\frac{1}{2 \kappa^{2}}\,,\nonumber\\
G_{5}(\phi, X)&=&0\,.
\end{eqnarray}
Eq.(\ref{first_friedmann_equation}), Eq.(\ref{second_friedmann_equation}) and Eq. (\ref{Klein-Gordan-equation}) can be rewritten by using Eq.(\ref{G_i_functions})  as,
\begin{eqnarray}
\frac{3H^{2}}{\kappa^{2}}&=& \frac{\dot{\phi}^{2}}{2}+V(\phi)-3\dot{\phi}^{3}HF(\phi)+\frac{\dot{\phi}^{4}}{2}F_{,\phi}+\rho_{m}+\rho_{r} \label{model_first_friedmann}\,,\\
-\frac{\dot{2H}}{\kappa^{2}}&=& \dot{\phi}^{2}+\dot{\phi}^{4}F_{,\phi}+\ddot{\phi}\dot{\phi}^{2}F(\phi)-3\dot{\phi}^{3} H F(\phi)+\rho_m+\rho_{r}+p_{r} \label{model_second_friedmann}\,,
\end{eqnarray}
\begin{eqnarray}\label{model_Klein_gordan}
&\ddot{\phi}+3H\dot{\phi}+2\ddot{\phi}\dot{\phi}^{2}F_{,\phi}+\frac{1}{2}\dot{\phi}^{4} F_{,\phi \phi}-3\dot{H}\dot{\phi}^{2}F(\phi)-6H\ddot{\phi}\dot{\phi}F(\phi)\nonumber \\ &-9H^{2}\dot{\phi}^{2}F(\phi)+V_{,\phi}=0\,.    
\end{eqnarray}
The Friedmann equations (\ref{model_first_friedmann}-\ref{model_second_friedmann}) can also be respectively represented as
\begin{eqnarray}
    \frac{3}{\kappa^{2}}H^{2}&=&\rho_{m}+\rho_{r}+\rho_{de}\,, \label{Einstein_first_friedmann}\\
    -\frac{2}{\kappa^{2}}\dot{H}&=&\rho_{m}+\frac{4}{3}\rho_{r}+\rho_{de}+p_{de}\,. \label{Einstein_second_friedmann}
\end{eqnarray}
Now, comparing Eq. (\ref{model_first_friedmann}) with Eq. (\ref{Einstein_first_friedmann}) and Eq.  (\ref{model_second_friedmann}) with Eq. (\ref{Einstein_second_friedmann}), the energy density ($\rho_{de}$) and pressure ($p_{de}$) for the dark energy sector can be determined as 

\begin{eqnarray}
\rho_{de}&=& \frac{\dot{\phi}^{2}}{2}+V(\phi)-3\dot{\phi}^{3}HF(\phi)+\frac{\dot{\phi}^{4}}{2}F_{,\phi} \label{rho_de}\,, \\ 
p_{de}&=& \frac{\dot{\phi}^{2}}{2}-V(\phi)+\ddot{\phi}\dot{\phi}^{2}F(\phi)+\frac{\dot{\phi}^{4}}{2}F_{,\phi}  \label{p_de}\,.
\end{eqnarray}
Furthermore, we can define the dark energy EoS parameter as
\begin{eqnarray}\label{w_de}
\omega_{de}=\frac{p_{de}}{\rho_{de}}= \frac{\frac{\dot{\phi}^{2}}{2}-V(\phi)+\ddot{\phi}\dot{\phi}^{2}F(\phi)+\frac{\dot{\phi}^{4}}{2}F_{,\phi}}{\frac{\dot{\phi}^{2}}{2}+V(\phi)-3\dot{\phi}^{3}HF(\phi)+\frac{\dot{\phi}^{4}}{2}F_{,\phi}}\,.    
\end{eqnarray}
One major benefit of Galileon cosmology is that $\omega_{de}$ can be quintessence-like or phantom-like or undergo the phantom divide crossing throughout the evolution according to $F(\phi)$. Moreover, Eq. (\ref{rho_de}) and Eq. (\ref{p_de}) satisfy the conservation equation
\begin{equation} \label{conser_rho_de}
\dot{\rho}_{de}+3H(\rho_{de}+p_{de})=0\,.   
\end{equation}
The conservation equation for matter and radiation sectors takes the standard form,
\begin{eqnarray}
\dot{\rho}_{m}+3H\rho_{m}=0\,, \\
\dot{\rho}_{r}+4H\rho_{r}=0\,.
\end{eqnarray}
The total EoS parameter is defined as 
\begin{equation}\label{w_tot}
\omega_{tot}=\frac{p_{r}+p_{de}}{\rho_{m}+\rho_{r}+\rho_{de}} =-1-\frac{2\dot{H}}{3H^{2}}\,.   
\end{equation}
The deceleration parameter can be written as 
\begin{equation}\label{qece}
q=-1-\frac{\dot{H}}{H^{2}} =\frac{1}{2}(1+3\omega_{tot})\,.  \end{equation}
By observation, we can infer that $q<0$ indicates acceleration, while $q>0$ implies the decelerating phase of the Universe. We shall now define the dynamical variables and autonomous systems to study the different phases and behaviors of the Universe through dynamical system analysis in Galileon cosmology.

\section{Dynamical System Framework}\label{SECIII}
The dynamical system approach is a potent technique that helps to explain the overall development and global dynamics of the Universe. A specific cosmological model may be expressed as an independent system of certain differential equations by carefully selecting the dynamical variables. This also provides a prompt response to whether the model can replicate the observed expansion of the Universe. First, we need to define the autonomous systems: $X^{'}=f(X)$, where $X$ is the column vector and prime ($'$) denotes the derivative for $N=lna$. Here, we propose the following dimensionless variables to generate the relevant autonomous system associated with the set of cosmological equations:
\begin{eqnarray}\label{dynamical_variable}
x=\frac{\kappa\dot{\phi}}{\sqrt{6}H}\,, \hspace{1cm} y=\frac{\kappa\sqrt{V}}{\sqrt{3}H}\,, \hspace{1cm} u=H\dot{\phi} F(\phi)\,, \hspace{1cm} \rho=\frac{\kappa \sqrt{\rho_{r}}}{\sqrt{3}H} \,,
\end{eqnarray}
and 
\begin{eqnarray}\label{dynamical_variable1}
\alpha=-\frac{F^{'}(\phi)}{\kappa F(\phi)}\,, \hspace{0.2cm} \lambda=-\frac{V^{'}(\phi)}{\kappa V(\phi)}\,, \hspace{0.2cm} \Theta=\frac{F^{''}(\phi)F(\phi)}{F^{'}(\phi)^{2}}\,, \hspace{0.2cm} \Gamma=\frac{V^{''}(\phi)V(\phi)}{V^{'}(\phi)^{2}}\,.
\end{eqnarray}
In term of dynamical system variables, the density parameters can be expressed as,
\begin{eqnarray}
\Omega_{de}&=(1-6u)x^{2}+y^{2}-\sqrt{6}x^{3}u\alpha\,, \label{density-parameter-dynamical variable1} \\ 
\Omega_{r}&=\rho^{2}\,, \label{density-parameter-dynamical variable2} \\ 
\Omega_{m}&=1-(1-6u)x^{2}-y^{2}+\sqrt{6}x^{3}u\alpha-\rho^{2}\,. \label{density-parameter-dynamical variable3}     
\end{eqnarray}
The constraint equation can be written as,
\begin{equation} \label{Constraint equation}
\Omega_{de}+\Omega_{m}+\Omega_{r}=1 .   
\end{equation}
In terms of dynamical variables, the background cosmological parameters can be defined as,
\begin{eqnarray}
 q&=\frac{1}{2 u \left(9 u x^2+2 \sqrt{6} \alpha  x+6\right)+2} \bigg[\rho ^2+u x \bigg(\sqrt{6} \bigg(\alpha  \left(2 \rho ^2+3 x^2-6 y^2+2\right)\nonumber\\ & +3 \lambda  y^2\bigg)  -18 u x \left(\alpha  x \left(\alpha  (\Theta +2) x+\sqrt{6}\right)-2\right)\bigg)\nonumber\\ & +6 u \left(\rho ^2-3 y^2+1\right)+3 x^2-3 y^2+1\bigg] \,,\label{qmodel} \\   
\omega_{tot}&=\frac{1}{3 u \left(9 u x^2+2 \sqrt{6} \alpha  x+6\right)+3}\bigg[-18 \alpha ^2 (\Theta +2) u^2 x^4+3 \left(9 u^2+1\right) x^2 \nonumber\\ & +3 \sqrt{6} \alpha  u (1-6 u) x^3 +\sqrt{6} u x \left(2 \alpha  \rho ^2+3 y^2 (\lambda -2 \alpha )\right)\nonumber\\ & -(6 u+1) \left(3 y^2-\rho ^2\right)\bigg] \,,\label{omegatotmodel} \\       
\omega_{de}&=\frac{1}{-\sqrt{6} \alpha  u x^3+(1-6 u) x^2+y^2} \bigg[\frac{\sqrt{6} \lambda  u^3 y^2}{x \left(2 \sqrt{6} \alpha  u x+6 u+1\right)}-\frac{6 \alpha ^2 \Theta  u^2 x^4}{2 \sqrt{6} \alpha  u x +6 u+1}\nonumber\\ & +\frac{18 u^2 x^2}{2 \sqrt{6} \alpha  u x+6 u+1} -\sqrt{6} \alpha  u x^3-\frac{6 u x^2}{2 \sqrt{6} \alpha  u x+6 u+1}+x^2-y^2\nonumber\\ & +\frac{1}{\left(2 \sqrt{6} \alpha  u x+6 u+1\right) \left(2 u \left(9 u x^2+2 \sqrt{6} \alpha  x+6\right)+2\right)} \bigg(6 u^2 x^2 \bigg(18 \alpha ^2 \nonumber\\ &(\Theta +2) u^2 x^4 -3 \left(18 u^2+1\right) x^2+3 \sqrt{6} \alpha  u (6 u-1) x^3+\nonumber\\ &\sqrt{6} u x \left(-2 \alpha  \left(\rho ^2+3\right)+6 \alpha  y^2-3 \lambda  y^2\right)\nonumber\\ &+(6 u+1) \left(-\rho ^2+3 y^2-3\right)\bigg)\bigg) \bigg] \label{omegademodel}.
\end{eqnarray}
The background parameters display the evolution of the Universe through the fixed points of the autonomous system. It includes various phases of the Universe, accelerating and decelerating phases, quintessence, phantom, and quintom eras of the Universe. To obtain an autonomous system, we need to take the derivative of Eq. (\ref{dynamical_variable}) and Eq. (\ref{dynamical_variable1}) with respect to $N$,

\begin{eqnarray}
\frac{dx}{dN}&=&\frac{\sqrt{6} \lambda  y^2-6 \left(\alpha ^2 \Theta  u x^3-3 u x+x\right)}{4 u \left(\sqrt{6} \alpha  x+3\right)+2}+\frac{x \left(\frac{3 u}{2 \sqrt{6} \alpha  u x+6 u+1}-1\right)}{2 u \left(9 u x^2+2 \sqrt{6} \alpha  x+6\right)+2} \nonumber \\ &&\bigg[18 \alpha ^2 (\Theta +2) u^2 x^4-3 \left(18 u^2+1\right) x^2 +3 \sqrt{6} \alpha  u (6 u-1) x^3 +\sqrt{6} u x \nonumber \\ &&\left(-2 \alpha  \left(\rho ^2+3\right)+6 \alpha  y^2-3 \lambda  y^2\right)+(6 u+1) \left(-\rho ^2+3 y^2-3\right)\bigg]\,, \label{autonomous-system1}\\
\frac{dy}{dN}&=& -\sqrt{\frac{3}{2}} \lambda  x y+\frac{1}{2 u \left(9 u x^2+2 \sqrt{6} \alpha  x+6\right)+2}\bigg[y \bigg(\rho ^2+u x \bigg(\sqrt{6} \bigg(\alpha  \bigg(2 \rho ^2 \nonumber \\&&+3 x^2-6 y^2+6\bigg) +3 \lambda  y^2\bigg) -18 u x \bigg(\alpha  x \left(\alpha  (\Theta +2)x+\sqrt{6}\right)-3\bigg)\bigg)\nonumber \\&&+6 u \left(\rho ^2-3 y^2+3\right)+3 x^2-3 y^2+3\bigg)\bigg]\,, \label{autonomous-system2} \\
\frac{du}{dN}&=& -\frac{3 \alpha ^2 \Theta  u^2 x^2}{2 \sqrt{6} \alpha  u x+6 u+1}+\frac{9 u^2}{2 \sqrt{6} \alpha  u x+6 u+1}-\sqrt{6} \alpha  u x -\frac{3 u}{2 \sqrt{6} \alpha  u x+6 u+1}\nonumber \\&&+\frac{\sqrt{\frac{3}{2}} \lambda  u y^2}{2 u x \left(\sqrt{6} \alpha  x+3\right)+x}+\frac{\frac{3 u^2}{2 \sqrt{6} \alpha  u x+6 u+1}+u}{2 u \left(9 u x^2+2 \sqrt{6} \alpha  x+6\right)+2} \bigg[18 \alpha ^2 (\Theta +2) u^2 x^4\nonumber \\&&-3 \left(18 u^2+1\right) x^2+3 \sqrt{6} \alpha  u (6 u-1) x^3 +\sqrt{6} u x \bigg(-2 \alpha  \left(\rho ^2+3\right) \nonumber \\&&+6 \alpha  y^2-3 \lambda  y^2\bigg)+(6 u+1) \left(-\rho ^2+3 y^2-3\right)\bigg]\,, \label{autonomous-system3} \\
\frac{d\rho}{dN}&=&\frac{1}{2 u \left(9 u x^2+2 \sqrt{6} \alpha  x+6\right)+2} \bigg[\rho  \bigg(\rho ^2+u x \bigg(\sqrt{6} \bigg(\alpha  \left(2 \rho ^2+3 x^2-6 y^2-2\right)\nonumber \\&&+3 \lambda  y^2\bigg)-18 u x \left(\alpha  x \left(\alpha  (\Theta +2) x+\sqrt{6}\right)-1\right)\bigg)\nonumber \\&&+6 u \left(\rho ^2-3 y^2-1\right)+3 x^2-3 y^2-1\bigg)\bigg]\,,\label{autonomous-system4} \\ 
\frac{d\alpha}{dN}&=& -\sqrt{6} \alpha ^2 (\Theta -1) x \,,\label{autonomous-system5} \\  
\frac{d\lambda}{dN}&=& -\sqrt{6} \lambda ^2 (\Gamma -1) x \,.\label{autonomous-system6}
\end{eqnarray} 
Dynamic systems (\ref{autonomous-system1}-\ref{autonomous-system6}) are not autonomous systems without knowing $\Gamma$ and $\Theta$ parameters. The parameters depend on the coupling scalar field $F(\phi)$ with the Galileon term and the potential scalar field $V(\phi)$. So, we need to define the functions $F(\phi)$ and $V(\phi)$ to get the autonomous systems. We have considered the exponential form of the coupling scalar field function $F(\phi)=F_{0}e^{-\alpha \kappa \phi(t)}$, where $\alpha$ is a dimensionless parameter. For this exponential form of the coupling scalar field, we obtain $\Theta=1$, which means $\frac{d\alpha}{dN}=0$. 
In addition, we have taken different forms of the potential function $V(\phi)$ to define the value of the parameter $\Gamma$. Other potential functions yield different forms of $\Gamma$, where $\Gamma$ depends on $\lambda$. Remember that the phase space analysis only applies to potentials where $\Gamma$ can be expressed as a function of $\lambda$. To close the system (\ref{autonomous-system1}-\ref{autonomous-system6}), we consider three well-motivated forms of $V(\phi)$ in three different cases and will provide a detailed analysis for the torsion Galileon theory.

\subsection{{\large Case-I:} $V(\phi)=V_{0}e^{-\lambda \kappa \phi(t)}$ \cite{Copeland_1998, Ng_2001}} \label{case1exp}

For exponential form of $V(\phi)$, $\Gamma=1$ and so $\frac{d\lambda}{dN}=0$. The dynamical system (\ref{autonomous-system1}-\ref{autonomous-system6}) reduces to four dimensions as well as the autonomous system. To obtain the critical points ($x_c, y_c, u_c, \rho_c$) of the autonomous system (\ref{autonomous-system1}-\ref{autonomous-system6}), we need to take the condition $\frac{dx}{dN}=\frac{dy}{dN}=\frac{du}{dN}=\frac{d\rho}{dN}=0$. After imposing the conditions, we have obtained the six critical points with their existence conditions as given in TABLE-\ref{TABLE-I}. According to cosmological observations \cite{SupernovaSearchTeam:1998fmf, SupernovaCosmologyProject:1998vns}, the phase of the Universe is accelerating expansion ($H>0$). Thus, the condition on the dynamical variables $x_c$, $y_c$, $u_c$ and $\rho_c$ must be real with $y_c>0$ and $\rho_c>0$. We have taken only positive values of $y_c$ since the negative value of $y_c$ describes the contraction phase ($H<0$) of the Universe. The corresponding value of the background cosmological parameters of each critical point has been displayed in TABLE-\ref{TABLE-II}. Finally, it should be noted from the physical viability condition [$0<\Omega_{i}<1$, where i= matter (m), radiation (r), dark energy (de)] of the energy density parameters that the physical condition $\rho_{r} \geq 0$ implies $\rho \geq 0$ and $(-\sqrt{6} \alpha u x^3 + (1-6u) x^2 + y^2) < 1$ and $y > 0$.

\begin{table}[H]
     \renewcommand{\arraystretch}{1.5} 
     \setlength{\tabcolsep}{8pt} 
    \caption{Critical points and existence condition {\bf (Case-I)}.  } 
    \centering 
    \begin{tabular}{|c|c|c|c|c|c|} 
    \hline\hline 
    C.P. & $x_{c}$ & $y_{c}$ & $u_{c}$ & $\rho_{c}$ & Exists for \\ [0.5ex] 
    \hline\hline 
    $A_{1}$  & $0$ & $0$ & $0$ & $0$ &$Always$ \\
    \hline
    $A_{2}$  & $0$ & $1$ & $0$ & $0$ &$\lambda=0$ \\
    \hline
    $A_{3 \pm}$  & $\pm1$ & $0$ & $0$ & $0$ &$Always$ \\
    \hline
    $A_{4}$  & $\frac{2 \sqrt{\frac{2}{3}}}{\lambda }$ & $\frac{2}{\sqrt{3} \lambda }$ & $0$ & $\frac{\sqrt{\lambda ^2-4}}{\lambda }$ &$\lambda \neq 0, \hspace{0.2cm} \lambda^2 \geq 4 $ \\
    \hline
    $A_{5}$  & $\frac{\sqrt{\frac{3}{2}}}{\lambda }$ & $\frac{\sqrt{\frac{3}{2}}}{\lambda }$ & $0$ & $0$ &$\lambda \neq 0$ \\
    \hline
    $A_{6}$  & $\frac{\lambda }{\sqrt{6}}$ & $\frac{\sqrt{6-\lambda ^2}}{\sqrt{6}}$ & $0$ & $0$ &$6 \geq \lambda^2 > 0$ \\
     [1ex] 
    \hline 
    \end{tabular}
    \label{TABLE-I}
\end{table}

\begin{table}[H]
     \renewcommand{\arraystretch}{2} 
     \setlength{\tabcolsep}{8pt} 
    \caption{Density parameters, Deceleration parameter, EoS parameters {\bf(Case-I}) } 
    \centering 
    \begin{tabular}{|c|c|c|c|c|c|c|} 
    \hline\hline 
    C.P. & $\Omega_{de}$ & $\Omega_{m}$ & $\Omega_{r}$ & $q$ & $\omega_{de}$ & $\omega_{tot}$ \\ [0.5ex] 
    \hline\hline 
    $A_{1}$  & $0$ & $1$ & $0$ & $\frac{1}{2}$ &$1$ & $0$\\
    \hline
    $A_{2}$  & $1$ & $0$ & $0$ & $-1$ &$-1$ & $-1$\\
    \hline
    $A_{3 \pm}$  & $1$ & $0$ & $0$ & $2$ &$1$ & $1$\\
    \hline
    $A_{4}$  & $\frac{4}{\lambda ^2}$ & $0$ & $1-\frac{4}{\lambda ^2}$ & $1$ &$\frac{1}{3}$ & $\frac{1}{3}$\\
    \hline
    $A_{5}$  & $\frac{3}{\lambda ^2}$ & $1-\frac{3}{\lambda ^2}$ & $0$ & $\frac{1}{2}$ &$0$ & $0$\\
    \hline
    $A_{6}$  & $1$ & $0$ & $0$ & $\frac{1}{2} \left(\lambda ^2-2\right)$ &$\frac{1}{3} \left(\lambda ^2-3\right)$& $\frac{1}{3} \left(\lambda ^2-3\right)$ \\
     [1ex] 
    \hline 
    \end{tabular}
    \label{TABLE-II}
\end{table}

{\bf \large {Description of Critical Points:}}
\begin{itemize}
\item Critical point $A_{1}$ provides matter-dominated solution $\Omega_{m}=1$ with total EoS parameter $\omega_{tot}=0$ and dark energy sector EoS parameter $\omega_{de}=1$. The positive value of the deceleration parameter ($q=\frac{1}{2}$) shows the decelerating phase of the Universe. It exists always.

\item Critical point $A_{2}$ represents dark energy-dominated state $\Omega_{de}=1$ with the value of the total and dark energy-dominated EoS parameter is $-1$. For this critical point, the negative value of the deceleration parameter indicates that the Universe is having accelerating behavior. It exists for $\lambda=0$.

\item The solution of the density parameter for the critical point $A_{3\pm}$ behaves $\Omega_{de}=1$. Simultaneously, the solution of the total and dark energy-dominated EoS parameter is $1$. This critical point shows the stiff-matter phase. It cannot explain the accelerated phase of the Universe since $q>0$. It exists always.

\item The solution of the density parameter for the critical point $A_{4}$ is $\Omega_{de}=\frac{4}{\lambda^{2}}$, $\Omega_{m}=0$, and $\Omega_{r}=1-\frac{4}{\lambda^{2}}$. The solution depends on the parameter $\lambda$, hence the scaling solution. This scaling solution reflects the radiation-dark energy-dominated era of the Universe. For this point, we have obtained $\omega_{tot}= \omega_{de}=\frac{1}{3}$. So, $A_{4}$ depicts a non-standard radiation-dominated epoch of the Universe with a negligible dark energy contribution. The positive value of the deceleration parameter indicates the decelerating era of the Universe. We have obtained the restriction $ \lambda^2  > 4$ on the model parameter imposed by the physical condition $0<\Omega_{de}^{r}<1$.

\item The scaling solution of the density parameters for the critical point $A_{5}$ is $\Omega_{de}=\frac{3}{\lambda^{2}}$, $\Omega_{m}=1-\frac{3}{\lambda^{2}}$, and $\Omega_{r}=0$ with the solution of the total and dark energy dominated EoS parameter is $0$. The point shows the non-standard matter-dominated phase of the Universe. This scaling solution represents the matter-dark energy-dominated era of the Universe. The positive value of the deceleration parameter indicates the decelerating phase. On imposing the condition $0<\Omega_{de}^{m}<1$, we obtain the constraint $\lambda^{2}>3$. It exists for $\lambda \neq 0$.

\item The critical point $A_{6}$ represented the dark energy-dominated phase of the Universe with $\Omega_{de}=1$ and $\omega_{tot}=\omega_{de}=-1+\frac{\lambda^2}{3}$. This critical point can explain the accelerated phase of the Universe for $\lambda^2 <2$. The condition $6\geq\lambda^{2}> 0$ indicates its existence. This critical point shows the quintessence phase ($-1<\omega_{tot}<-\frac{1}{3}$) of the Universe for the conditions $-\sqrt{2}<\lambda <0 $ and $0<\lambda <\sqrt{2}$ and shows the Phantom phase ($\omega_{tot}<-1$) for $\lambda <-\sqrt{6}$ and $\lambda >\sqrt{6}$. 
\end{itemize} 
{\bf \large{Stability Analysis:}}\\
Small perturbations around the critical points have been introduced to assess the stability of critical points, and the equations of the system have been linearized. So, we can determine the stability of the matrix $\mathcal{M}$ by determining its eigenvalues $\mu_1$, $\mu_2$, $\mu_3$, and $\mu_4$. The stability features are classified as (a) stable node; all the eigenvalues are negative; (b) unstable node, all the eigenvalues are positive; (iii) saddle point, in the presence of both positive and negative eigenvalues; and (iv) stable spiral: complex with negative real part eigenvalues. Stable nodes and stable spirals describe the late-time cosmic acceleration phase of the Universe without depending on the initial conditions. The eigenvalues of the Jacobean matrix and stability conditions for each critical point are listed below.

\begin{figure}[H]
 \centering
 \includegraphics[width=100mm]{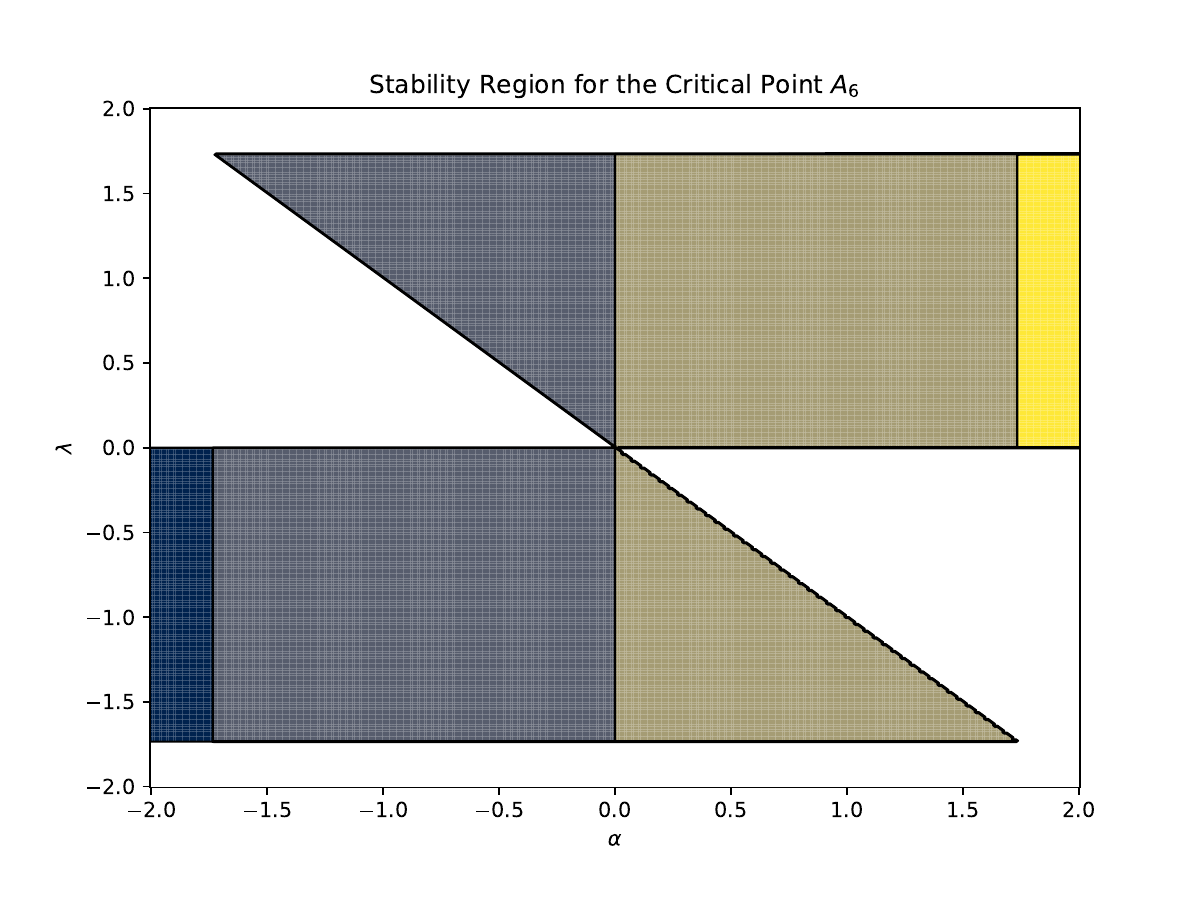}
 \caption{Stability region of the critical point $A_{6}$ between the model parameters $\alpha$ and $\lambda$.} \label{FigA6}
 \end{figure}

\begin{itemize}
\item  Eigenvalues of critical point $A_{1}$
\begin{eqnarray*}
\mu_{1} = -\frac{1}{2}, \hspace{0.2cm} \mu_{2} = -\frac{3}{2}, \hspace{0.2cm} \mu_{3} = -\frac{9}{2} , \hspace{0.2cm} \mu_{4} = \frac{3}{2} \,.     
\end{eqnarray*}
This point exhibits saddle behavior according to the linear stability theory as both positive and negative eigenvalues are present. 

\item Eigenvalues of critical point $A_{2}$
\begin{eqnarray*}
\mu_{1} = -2, \hspace{0.2cm} \mu_{2} = -3, \hspace{0.2cm} \mu_{3} = -3 , \hspace{0.2cm} \mu_{4} = -3 \,.   
\end{eqnarray*}
This point shows stable node behavior since all the eigenvalues are negative. This point can explain the accelerated phase of the Universe.

\item Eigenvalues of critical point $A_{3\pm}$
\begin{eqnarray*}
\mu_{1} = 3, \hspace{0.2cm} \mu_{2} = 1, \hspace{0.2cm} \mu_{3} = -6\pm\sqrt{6} \alpha , \hspace{0.2cm} \mu_{4} = 3\pm\sqrt{\frac{3}{2}} \lambda \,.   
\end{eqnarray*}

Sign ($-$) in $\mu_3$ and $\mu_{4}$ denote the eigenvalues of the critical point $A_{3+}$, whereas sign ($+$) describes the eigenvalues of the critical point $A_{3-}$. For $\alpha > -\sqrt{6}$ and $\lambda >\sqrt{6}$, the saddle behaviour is represented by the point $A_{3+}$. For the conditions $\alpha <\sqrt{6}$ and $\lambda <-\sqrt{6}$, the saddle behaviour is described by the point $A_{3-}$. The behavior of the unstable (node) if one of these two key points fails to satisfy the requirements mentioned above. 
    
\item Eigenvalues of critical point $A_{4}$
\begin{eqnarray*}
&\mu_{1} = 1, \hspace{0.2cm} \mu_{2} = -\frac{4 (\alpha +\lambda )}{\lambda }, \hspace{0.2cm} \mu_{3} =-\frac{1}{2} -\frac{\sqrt{64 \lambda ^6-15 \lambda ^8}}{2 \lambda ^4}, \nonumber \\& \mu_{4} = -\frac{1}{2} +\frac{\sqrt{64 \lambda ^6-15 \lambda ^8}}{2 \lambda ^4} \,.   
\end{eqnarray*}

This critical point shows unstable (saddle) behavior for the conditions $\bigg(-\frac{8}{\sqrt{15}}\leq \lambda <-2\land \alpha <-\lambda\bigg)$ and $\bigg(2<\lambda  \leq \frac{8}{\sqrt{15}}\land \alpha >-\lambda\bigg)$. Failing to satisfy the above conditions, the point exhibits unstable (node) behavior.

\item Eigenvalues of critical point $A_{5}$
\begin{eqnarray*}
&\mu_{1} = -\frac{1}{2}, \hspace{0.2cm} \mu_{2} = -\frac{3 (\alpha +\lambda )}{\lambda }, \hspace{0.2cm} \mu_{3} = -\frac{3 \left(\lambda ^4+\sqrt{24 \lambda ^6-7 \lambda ^8}\right)}{4 \lambda ^4}, \nonumber \\& \mu_{4} = -\frac{3}{4}+\frac{3 \sqrt{24 \lambda ^6-7 \lambda ^8}}{4 \lambda ^4} \,. 
\end{eqnarray*}
This critical point describes the stable node behavior for the conditions  
$\left(-2 \sqrt{\frac{6}{7}}\leq \lambda <-\sqrt{3}\land \alpha <-\lambda \right)$ and $\left(\sqrt{3}<\lambda \leq 2 \sqrt{\frac{6}{7}}\land \alpha >-\lambda \right)$. The point exhibits saddle behavior if it does not fulfill the above-mentioned conditions. 

\item Eigenvalues of critical point $A_{6}$
\begin{eqnarray*}
\mu_{1} = \frac{1}{2} \left(\lambda ^2-6\right), \hspace{0.2cm} \mu_{2} = \frac{1}{2} \left(\lambda ^2-4\right), \hspace{0.2cm} \mu_{3} =\lambda ^2-3 , \hspace{0.2cm}  \mu_{4} = -\lambda  (\alpha +\lambda ) \,.
\end{eqnarray*}
This critical point indicates stable node behavior for the conditions $\bigg(\alpha \leq -\sqrt{3}\land -\sqrt{3}<\lambda <0\bigg) $,  $\bigg(-\sqrt{3}<\alpha \leq 0\land \bigg(-\sqrt{3}<\lambda <0\lor -\alpha <\lambda <\sqrt{3}\bigg)\bigg) $,$ \bigg(0<\alpha <\sqrt{3}\land \left(-\sqrt{3}<\lambda <-\alpha \lor 0<\lambda <\sqrt{3}\right)\bigg)$, and $ \bigg(\alpha \geq \sqrt{3}\land 0<\lambda <\sqrt{3}\bigg)$.  If the point fails to satisfy the above conditions, it behaves as an unstable node or saddle. For this critical point, we have displayed a stability region [FIG.-\ref{FigA6}] between the model parameters $\alpha$ and $\lambda$ for the above-mentioned stable node conditions. The shaded region in  FIG.-\ref{FigA6} shows the stable node behavior. Late-time cosmic acceleration can be studied through this critical point.
\end{itemize}

{\bf { \large Numerical Solutions:}}

According to the solution of background parameters, the critical points represent different phases of the Universe. Among these critical points, $A_{2}$ and $A_{6}$ represent the dark energy-dominated phase of the Universe. Now, we wish to analyze the autonomous system (\ref{autonomous-system1}-\ref{autonomous-system6}) with the numerical solution. To obtain this, we used the NDSolve command in Mathematica. In addition, we compared our results with Hubble and Supernovae Ia (SNe Ia) observational data sets [Details in section--\ref{dataset}]. \\

\begin{figure}
     \centering
     \begin{subfigure}[b]{0.49\textwidth}
         \centering
         \includegraphics[width=\linewidth]{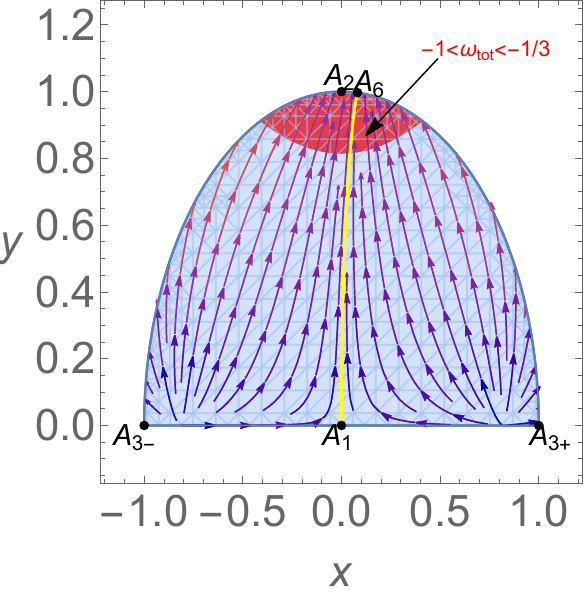}
         \caption{Phase space with $\alpha=-5$ and $\lambda=0.2$}
         \label{fig:case1_phaseportrait}
     \end{subfigure}
     \hfill
     \begin{subfigure}[b]{0.49\textwidth}
         \centering
         \includegraphics[width=\linewidth]{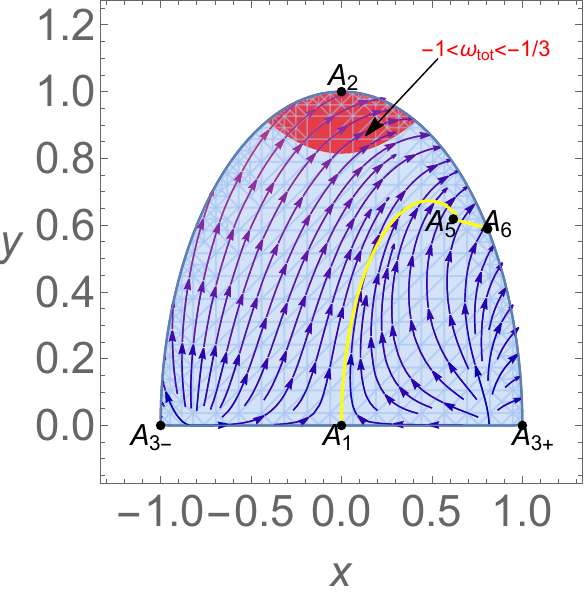}
         \caption{Phase space with $\alpha=-5$ and $\lambda=1.98$}
         \label{fig:case1_phaseA5}
     \end{subfigure}
     \hfill
     \begin{subfigure}[b]{0.49\textwidth}
         \centering
         \includegraphics[width=\linewidth]{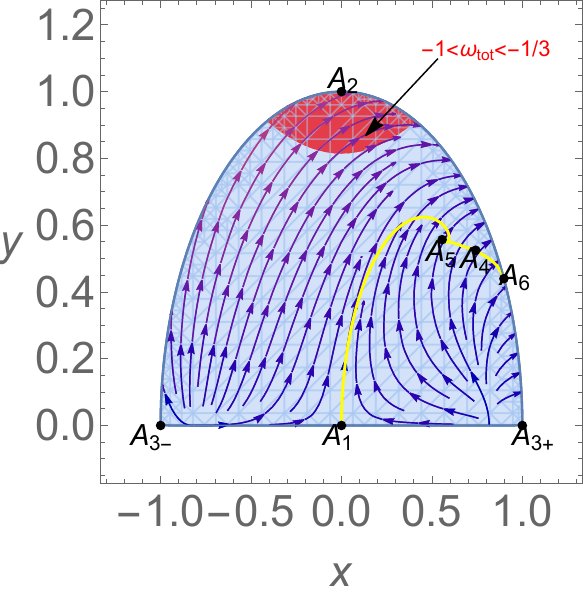}
         \caption{Phase space with $\alpha=-5$ and $\lambda=2.2$}
         \label{fig:case1_phaseA4}
     \end{subfigure}
\caption{Case1: 2D phase space portrait of the autonomous system (\ref{autonomous-system1}-\ref{autonomous-system6}). The red/shaded region indicates where the universe experiences accelerated expansion ($-1 < \omega_{tot} < -\frac{1}{3}$). The initial conditions are set to: $x = 10^{-5}$, $y = 9 \times 10^{-13}$, $u = 10^{-5}$, and $\rho = \sqrt{0.999661}$.} 
\label{Fig2phasepor}
\end{figure}

In FIG.~\ref{Fig2phasepor}, we analyze the phase space portrait for different model parameter values $\lambda$. The behavior within the phase space is categorized into three distinct regions, each corresponding to specific ranges of $\lambda$. This variation in $\lambda$ is crucial as it underscores the existence of critical points and delineates the accelerating region, represented in red/shaded areas, which includes the critical points $A_4$, $A_5$, and $A_6$. The critical points within this accelerating region are significant for understanding the dynamics of the Universe, as they indicate the conditions under which cosmic acceleration occurs. \\

In FIG.~\ref{fig:case1_phaseportrait}, we illustrate the phase portrait for $\lambda = 0.2$. At this value, the critical points $A_4$ and $A_5$ are absent as these points arise only when $\lambda^2 > 4$ and $\lambda^2 > 3$, respectively. These conditions adhere to the physical constraints on the density parameters $(0 < \Omega_r < 1)$ and $(0 < \Omega_m < 1)$. Consequently, in the accelerating region associated with the critical point $A_6$, where $\lambda^2 < 2$, the critical points $A_4$ and $A_5$ do not emerge. For $\lambda = 0.2$, the critical point resides in the accelerating region (red/shaded), with $A_6$ corresponding to the accelerating phase of the Universe under the condition $\lambda^2 < 2$. Thus, we have selected $\lambda = 0.2$ since it meets the stability criteria for the critical point $A_6$. All trajectories in the phase space consist of heteroclinic orbits that start from the points $A_{3\pm}$ and end at $A_6$. Two heteroclinic orbits are observed: $A_{3\pm} \to A_1 \to A_6$. These orbits can serve as physical models for the transition from dark matter to dark energy, effectively capturing the late-time evolution of the Universe, with the total EoS parameter, $\omega_{tot} = -1 + \frac{\lambda^2}{3}$. However, at early times, the model predicts stiff fluid domination represented by the points $A_{3\pm}$, which is unfavorable from a phenomenological viewpoint. For values of $\lambda^2 > 2$, the point $A_6$ would fall outside the acceleration region (red/shaded) and would not represent an inflationary solution. At $\lambda=0.2$, the critical point $A_6$ exhibits stable node behavior, indicating late-time cosmic acceleration of the Universe. The heteroclinic orbit solution (yellow line) is derived from the numerical solution of the autonomous system (\ref{autonomous-system1}-\ref{autonomous-system6}) with initial conditions $x = 10^{-5}$, $y = 9 \times 10^{-13}$, $u = 10^{-5}$ and $\rho = \sqrt{0.999661}$.\\

In FIG.-\ref{fig:case1_phaseA5}, we can see six critical points in the phase space for the condition $\lambda^2 > 3$. For $\lambda=1.98$, the critical point $A_4$ does not exist, and the critical point $A_6$ lies outside the accelerating region (red/shaded). So, the critical points $A_5$ and $A_6$ show saddle behavior (unstable) and indicate the decelerating phase. There are also two heteroclinic orbits $A_{3\pm} \to A_1 \to A_6$.\\

In FIG. \ref{fig:case1_phaseA4}, we observe seven critical points in the phase space for the condition $\lambda^2 > 4$. For $\lambda = 2.2$, the critical point $A_6$ is located outside the accelerating region (red/shaded). At this value, the critical points $A_4$, $A_5$, and $A_6$ exhibit saddle-like (unstable) behavior and show the decelerating phase. Here also, $A_6$ remains outside the acceleration region (red/shaded) and hence never corresponds to an inflationary solution. Two heteroclinic orbits exist: $A_{3\pm} \to A_1 \to A_6$. \\

\begin{figure}
     \centering
     \begin{subfigure}[b]{0.4\textwidth}
         \centering
         \includegraphics[width=\linewidth]{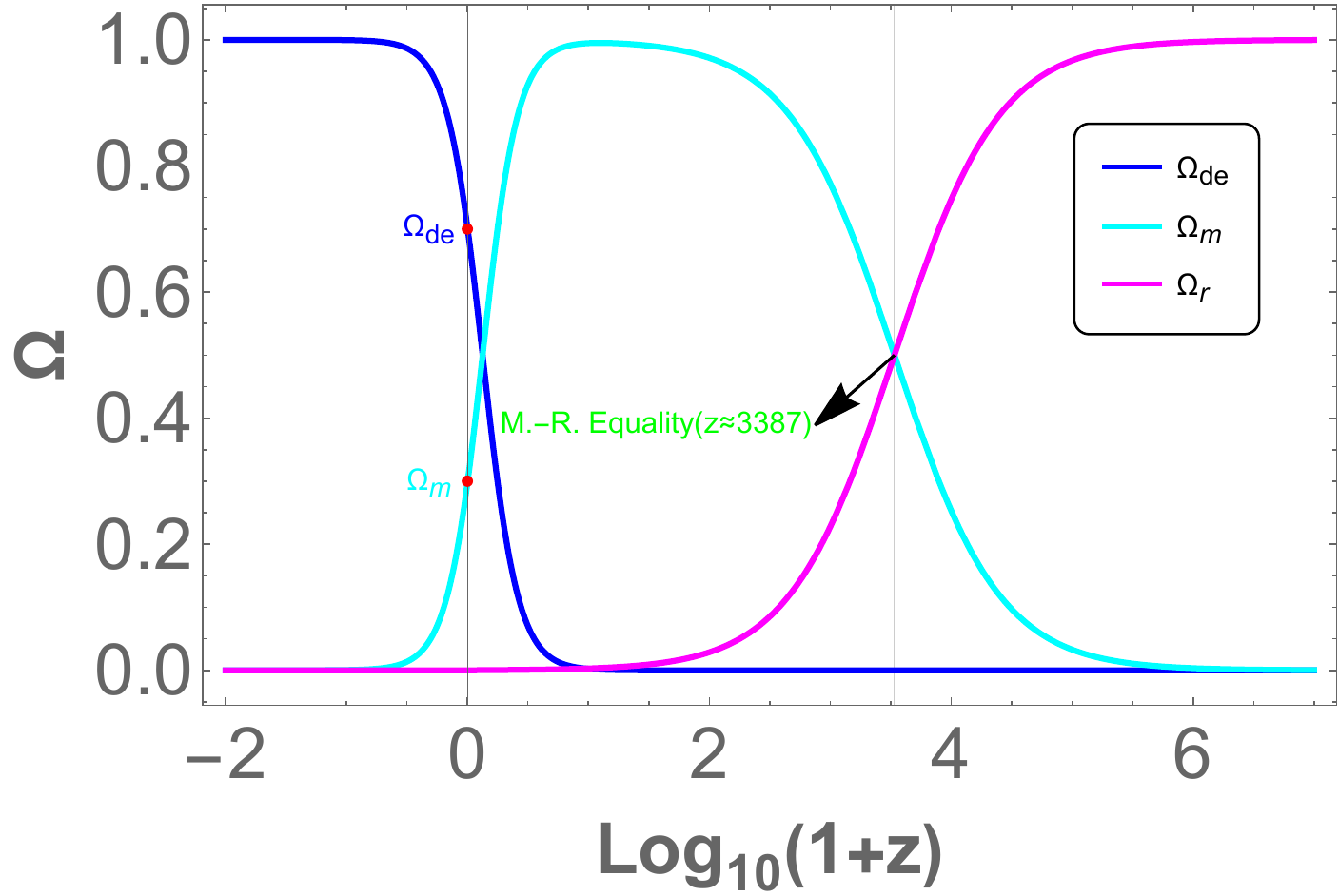}
         \caption{Evolution of density parameters.}
         \label{fig:case1_density}
     \end{subfigure}
     \hfill
     \begin{subfigure}[b]{0.4\textwidth}
         \centering
         \includegraphics[width=\linewidth]{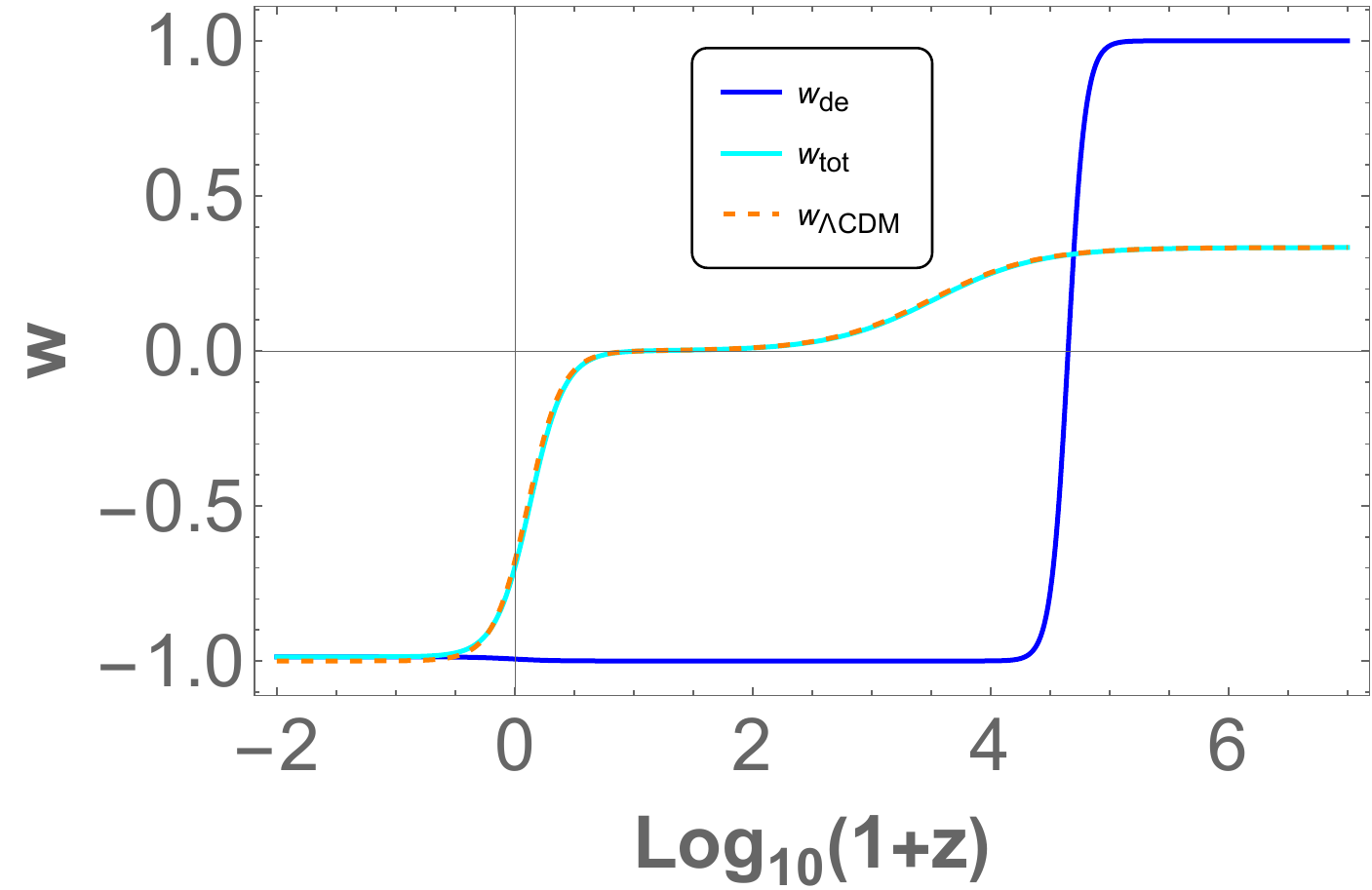}
         \caption{Evolution of EoS parameters in redshift.}
         \label{fig:case1_statepara}
     \end{subfigure}
\caption{In this figure, we set $\alpha=-5$ and $\lambda=0.2$ with the initial conditions are the same as in FIG.-\ref{Fig2phasepor}. The red dot indicates the present-day value of the density parameter at redshift $z = 0$, while the vertical black line corresponds to the current cosmological time. M.R. indicates the time of matter-radiation equality.} 
\label{Fig1}
\end{figure}

\begin{figure}
     \centering
     \begin{subfigure}[b]{0.4\textwidth}
         \centering
         \includegraphics[width=\linewidth]{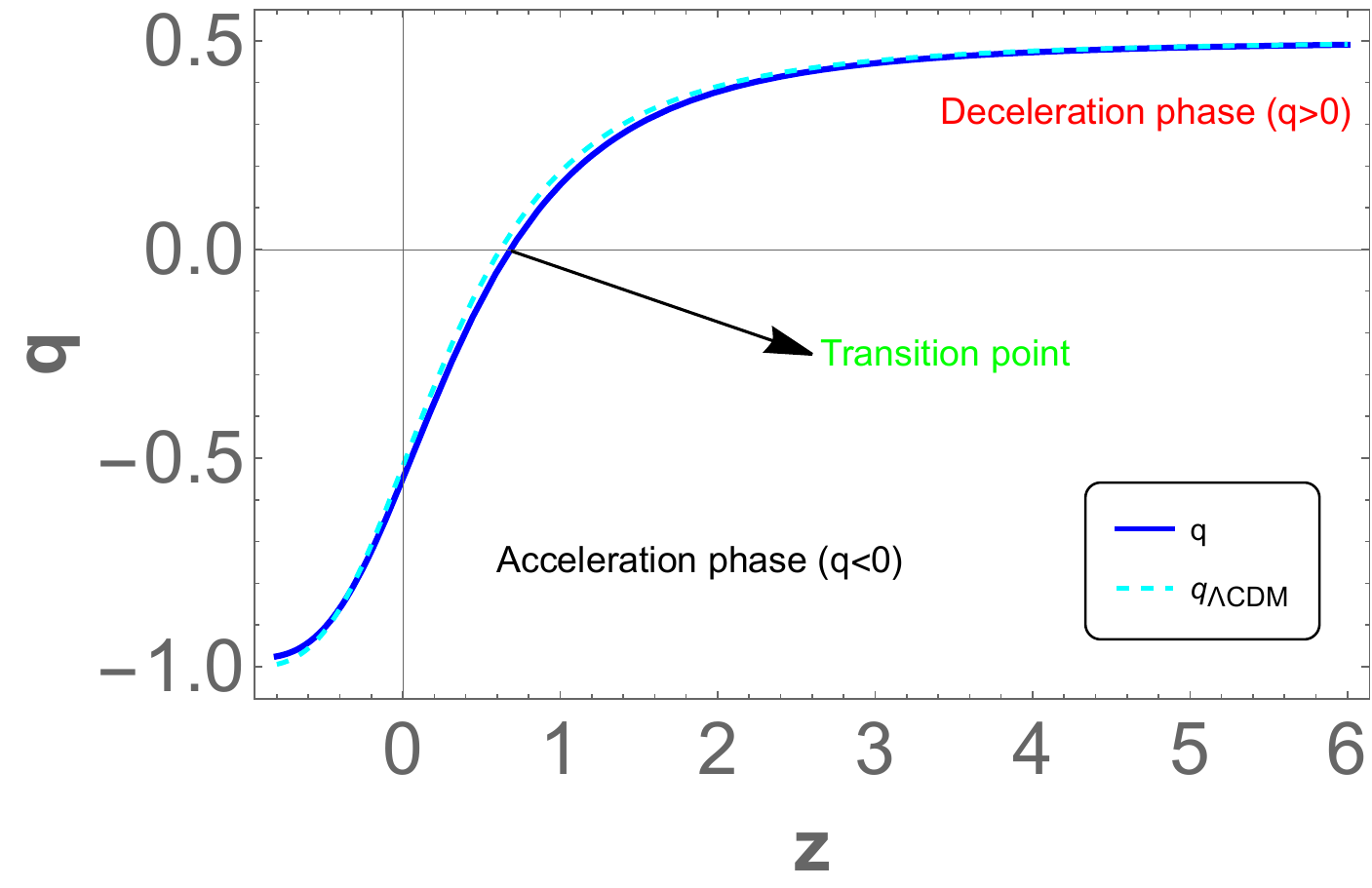}
         \caption{Evolution of deceleration parameter $q$ in redshift.}
         \label{fig:case1_qz}
     \end{subfigure}
     \hfill
     \begin{subfigure}[b]{0.4\textwidth}
         \centering
         \includegraphics[width=\linewidth]{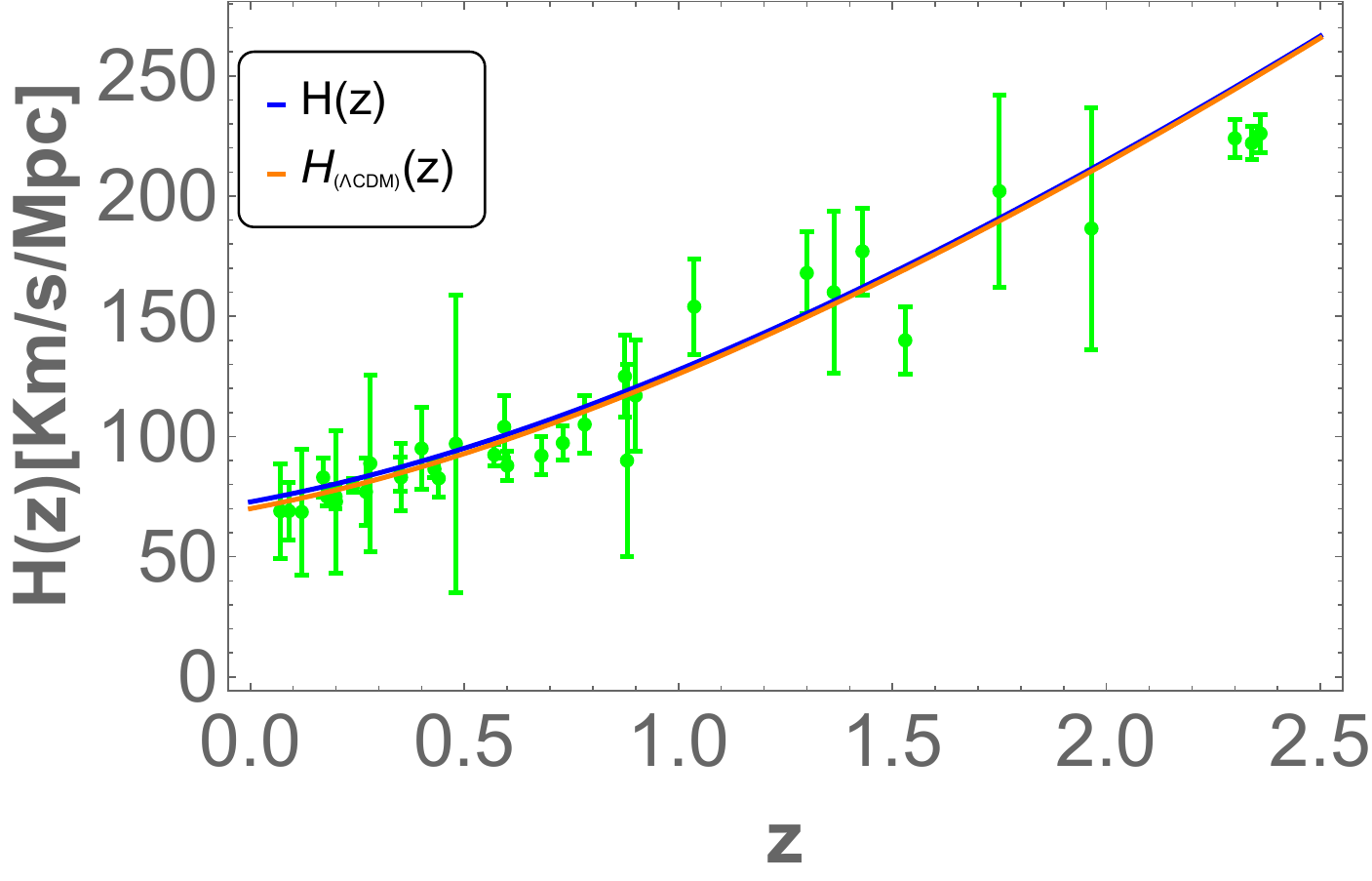}
         \caption{Evolution of Hubble rate $H(z)$ in redshift.}
         \label{fig:case1_Hz}
     \end{subfigure}
\caption{In this figure, we set $\alpha=-5$ and $\lambda=0.2$ with the initial conditions are the same as in FIG.-\ref{Fig2phasepor}.} 
\label{Fig2}
\end{figure}

\begin{figure}[H]
 \centering
 \includegraphics[width=80mm]{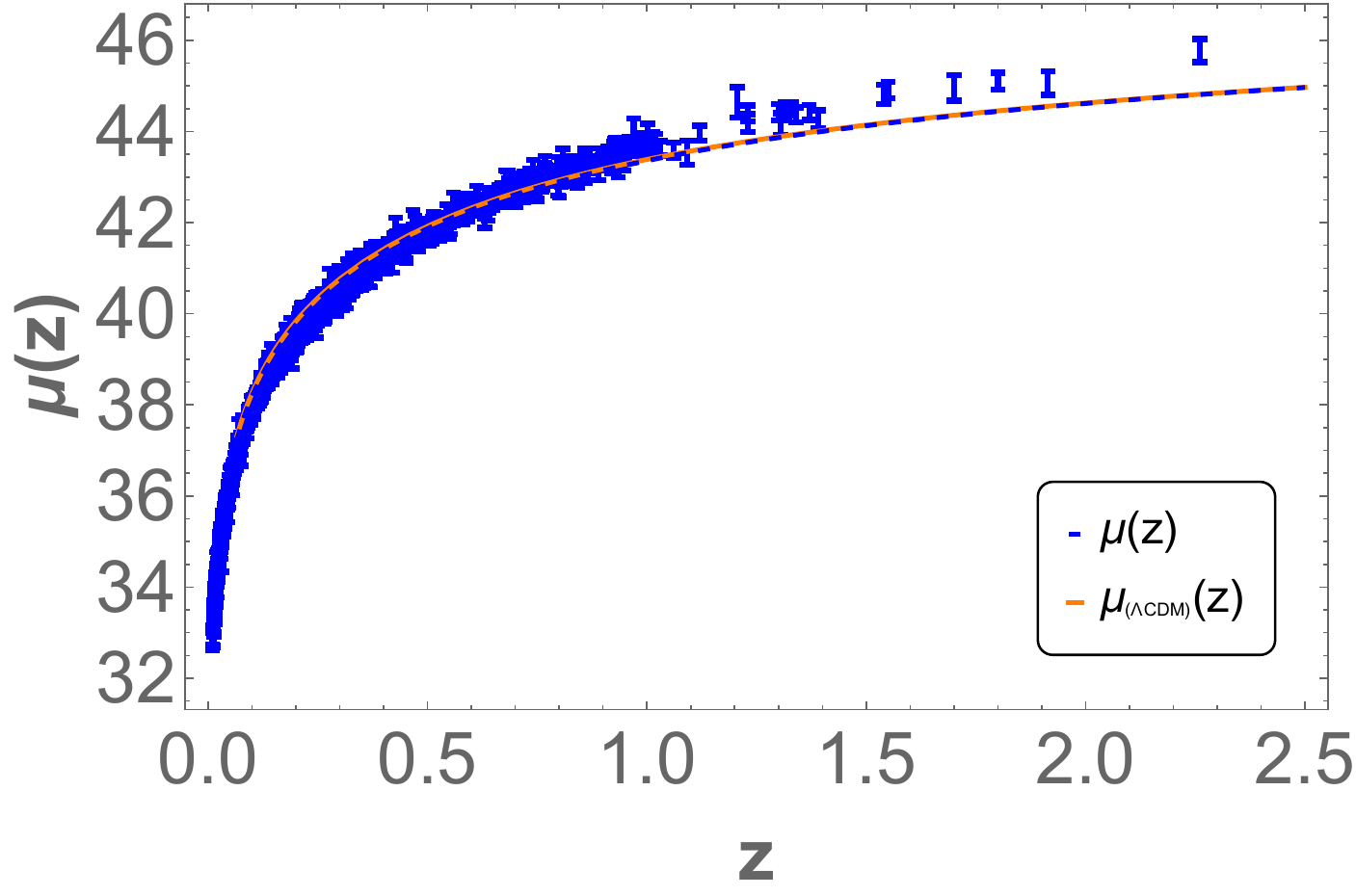} 
 \caption{In this figure, we set $\alpha=-5$ and $\lambda=0.2$ with the initial conditions are the same as in FIG.-\ref{Fig2phasepor}. The figure shows the evolution of distance modulus function $\mu(z)$ ( dashed blue) and $\Lambda$CDM model distance modulus function $\mu_{\Lambda CDM}(z)$ along with 1048 Supernovae Ia (SNe Ia)  data points \cite{Scolnic_2018}. } \label{Fig2pan}
 \end{figure}

From a physics perspective, the cosmological dynamics of the exponential potential are intriguing due to the emergence of late-time accelerated solutions that can be utilized to model dark energy. For these solutions to be cosmologically viable, a sufficiently flat potential ($\lambda^2<2$) is necessary, along with a strong fine-tuning of initial conditions to ensure dark energy domination persists. This solution must resemble the sequence:$A_{3\pm} \to A_1 \to A_6$ (FIG.-\ref{fig:case1_phaseportrait}). At early times, the only feasible solutions are the non-physical stiff fluid Universe. We have plotted the background cosmological parameters such as the  EoS parameters, energy densities, deceleration parameters, Hubble rate, and modulus function for a solution shadowing the heteroclinic sequence $A_{3\pm} \to A_1 \to A_6$ shown in FIG.-\ref{Fig1}, FIG.-\ref{Fig2} and FIG.- \ref{Fig2pan}.\\

Graphically, the relative energy densities of radiation, dark energy, and dark matter are shown in FIG.-\ref{fig:case1_density}. Radiation occurs in the early Universe, followed by a brief period of dominance over dark matter and, finally, the cosmological constant. The matter and dark energy sector density parameters are currently $\Omega_{m}\approx 0.3$ and $\Omega_{de}\approx 0.7$, respectively. The time of matter-radiation equality is around $z\approx 3387$. In FIG.-\ref{fig:case1_statepara}, the behaviour of EoS parameter shows that  $\omega_{tot}$ (cyan) starts from the radiation value of $\frac{1}{3}$, drops to $0$ during the matter-dominated era, and eventually reaches $-1$. Also, both $\omega_{de}$ (blue) and $\omega_{\Lambda CDM}$(dashed orange) approaches $-1$ at late-time. At present, $\omega_{de}(z=0)=-0.99$, which is compatible with the present Planck Collaboration result [$\omega_{de}(z=0)= -1.028 \pm 0.032$ \cite{Aghanim:2018eyx}]. In FIG.-\ref{fig:case1_qz}, the evolutionary behavior of the deceleration parameter shows transition behavior at $z\approx 0.66$, compatible with the current observational data \cite{PhysRevD.90.044016a}. The present value of the deceleration parameter is $q(z=0) \approx -0.55$, consistent with the visualized cosmological observations \cite{PhysRevResearch.2.013028}. In FIG.-\ref{fig:case1_Hz}, we illustrate the  Hubble rate evolution with the Hubble rate $H_{\Lambda CDM}(z)$, and the 31 Hubble data points \cite{Moresco_2022_25},$H_{0}=71.88$ Km/(Mpc sec) \cite{Aghanim:2018eyx}. We find our model to be fairly close to the standard $\Lambda$CDM model. In FIG.-\ref{Fig2pan}, we plot the evolution of the modulus function $\mu(z)$ and observe that the model curve and the $\Lambda$CDM model modulus function $\mu_{\Lambda CDM}$ are well within error bars. \\


\subsection{{\large Case-II:} $V(\phi)=V_{0}\phi^{n}$ \cite{Copeland:2006wr}} \label{case2power}

To find the value of $\Gamma$ in the power-law form of potential function $V(\phi)=V_{0}\phi^{n}$, we consider the function  $f(\lambda)=\lambda^{2}(\Gamma-1)$ [Eq. (\ref{autonomous-system6})]. The general structure of the function $f(\lambda)$ becomes $f(\lambda)=\beta_1 \lambda^2+ \beta_{2} \lambda+ \beta_{3}$. Here $\lambda^\pm=\frac{1}{2\beta_{1}}\left(-\beta_{2}\pm \sqrt{\beta_2^{2}-4\beta_{1} \beta_{3}}\right)$ are two roots of $f(\lambda)$. These roots always satisfy the condition $(\beta_2^{2}-4\beta_{1} \beta_{3}) \geq 0$ since the interest is only the real solutions of $f(\lambda)$. The autonomous system is reduced to five-dimension space for this form of $V(\phi)$. Now, we need to impose the conditions, $\frac{dx}{dN}=\frac{dy}{dN}=\frac{du}{dN}=\frac{d\rho}{dN}=\frac{d\lambda}{dN}=0$ to determine the critical points of the autonomous system (\ref{autonomous-system1}-\ref{autonomous-system6}). In TABLE-\ref{TABLE-III}, the eigenvalues of the critical points are presented along with the existence condition, whereas the values of the cosmological parameters are given in TABLE-\ref{TABLE-IV}. Now, the roots of the potential function $V(\phi)=V_{0}\phi^{n}$ are as follows:
\begin{eqnarray}\label{powerlawpotentialroot}
f(\lambda)&=&-\frac{\lambda^{2}}{2}\,, \hspace{0.5cm} \beta_{1}=-\frac{1}{n}\,, \hspace{0.5cm} \beta_{2}=\beta_{3}=0\,. \nonumber \\&&
\lambda^{+}=\lambda^{-}=-\frac{n}{2}
\end{eqnarray}
Since the roots $\lambda^{+}$ and $\lambda^{-}$ are same, therefore the behaviour of critical points $B_{3\pm}$, $B_{5}$, $B_{7}$ and $B_{9}$ respectively same as that of $B_{4\pm}$, $B_{6}$, $B_{8}$ and $B_{10}$.   
\begin{table}[H]
     \renewcommand{\arraystretch}{1.3} 
     \setlength{\tabcolsep}{7.5pt} 
    \caption{Critical points and existence condition.} 
    \centering 
    \begin{tabular}{|c|c|c|c|c|c|c|} 
    \hline\hline 
    C.P. & $x_{c}$ & $y_{c}$ & $u_{c}$ & $\rho_{c}$&$\lambda_c$ & Exists for \\ [0.5ex] 
    \hline\hline 
    $B_{1}$  & $0$ & $0$ & $0$ & $0$ &$\lambda_c$&$Always$ \\
    \hline
    $B_{2}$  & $0$ & $1$ & $0$ & $0$& $0$ &$\lambda=0$ \\
    \hline
    $B_{3 \pm}$  & $\pm1$ & $0$ & $0$ & $0$&$\lambda^{+}$ &$Always$ \\
    \hline
    $B_{4 \pm}$  & $\pm1$ & $0$ & $0$ & $0$&$\lambda^{-}$ &$Always$ \\
    \hline
    $B_{5}$  & $\frac{\sqrt{\frac{3}{2}}}{\lambda^{+} }$ & $\frac{\sqrt{\frac{3}{2}}}{\lambda^{+} }$ & $0$ & $0$ &$\lambda^{+}$&$\lambda^{+} \neq 0$ \\
      \hline
    $B_{6}$  & $\frac{\sqrt{\frac{3}{2}}}{\lambda^{-} }$ & $\frac{\sqrt{\frac{3}{2}}}{\lambda^{-} }$ & $0$ & $0$ &$\lambda^{-}$&$\lambda^{-} \neq 0$ \\
    \hline
    $B_{7}$  & $\frac{2 \sqrt{\frac{2}{3}}}{\lambda^{+} }$ & $\frac{2}{\sqrt{3} \lambda^{+} }$ & $0$ & $\frac{\sqrt{(\lambda^{+}) ^2-4}}{\lambda^{+} }$ &$\lambda^{+}$&$\lambda^{+} \neq 0, \hspace{0.2cm} (\lambda^{+})^2 \geq 4 $ \\
    \hline
    $B_{8}$  & $\frac{2 \sqrt{\frac{2}{3}}}{\lambda^{-} }$ & $\frac{2}{\sqrt{3} \lambda^{-} }$ & $0$ & $\frac{\sqrt{(\lambda^{-}) ^2-4}}{\lambda^{-} }$ &$\lambda^{-}$&$\lambda^{-} \neq 0, \hspace{0.2cm} (\lambda^{-})^2 \geq 4 $ \\
    \hline
    $B_{9}$  & $\frac{\lambda^{+} }{\sqrt{6}}$ & $\frac{\sqrt{6-(\lambda^{+} )^2}}{\sqrt{6}}$ & $0$ & $0$&$\lambda^{+}$ &$6 \geq (\lambda^{+})^2 > 0$ \\
      \hline
    $B_{10}$  & $\frac{\lambda^{-} }{\sqrt{6}}$ & $\frac{\sqrt{6-(\lambda^{-} )^2}}{\sqrt{6}}$ & $0$ & $0$&$\lambda^{-}$ &$6 \geq (\lambda^{-})^2 > 0$ \\
     [1ex] 
    \hline 
    \end{tabular}
    \label{TABLE-III}
\end{table}

\begin{table}[H]
     \renewcommand{\arraystretch}{2} 
     \setlength{\tabcolsep}{7.5pt} 
    \caption{Density parameters, Deceleration parameter, EoS parameters.} 
    \centering 
    \begin{tabular}{|c|c|c|c|c|c|c|} 
    \hline\hline 
    C.P. & $\Omega_{de}$ & $\Omega_{m}$ & $\Omega_{r}$ & $q$ & $\omega_{de}$ & $\omega_{tot}$ \\ [0.5ex] 
    \hline\hline 
    $B_{1}$  & $0$ & $1$ & $0$ & $\frac{1}{2}$ &$1$ & $0$\\
    \hline
    $B_{2}$  & $1$ & $0$ & $0$ & $-1$ &$-1$ & $-1$\\
    \hline
    $B_{3 \pm}$  & $1$ & $0$ & $0$ & $2$ &$1$ & $1$\\
    \hline
    $B_{4 \pm}$  & $1$ & $0$ & $0$ & $2$ &$1$ & $1$\\
    \hline
    $B_{5}$  & $\frac{3}{(\lambda^{+}) ^2}$ & $1-\frac{3}{(\lambda^{+}) ^2}$ & $0$ & $\frac{1}{2}$ &$0$ & $0$\\
     \hline
    $B_{6}$  & $\frac{3}{(\lambda^{-}) ^2}$ & $1-\frac{3}{(\lambda^{-}) ^2}$ & $0$ & $\frac{1}{2}$ &$0$ & $0$\\
    \hline
    $B_{7}$  & $\frac{4}{(\lambda^{+}) ^2}$ & $0$ & $1-\frac{4}{(\lambda^{+}) ^2}$ & $1$ &$\frac{1}{3}$ & $\frac{1}{3}$\\
    \hline
    $B_{8}$  & $\frac{4}{(\lambda^{-}) ^2}$ & $0$ & $1-\frac{4}{(\lambda^{-}) ^2}$ & $1$ &$\frac{1}{3}$ & $\frac{1}{3}$\\
    \hline
    $B_{9}$  & $1$ & $0$ & $0$ & $\frac{1}{2} \left((\lambda^{+}) ^2-2\right)$ &$\frac{1}{3} \left((\lambda^{+} )^2-3\right)$& $\frac{1}{3} \left((\lambda^{+}) ^2-3\right)$ \\
     \hline
    $B_{10}$  & $1$ & $0$ & $0$ & $\frac{1}{2} \left((\lambda^{-}) ^2-2\right)$ &$\frac{1}{3} \left((\lambda^{-} )^2-3\right)$& $\frac{1}{3} \left((\lambda^{-}) ^2-3\right)$ \\
     [1ex] 
    \hline 
    \end{tabular}
    \label{TABLE-IV}
\end{table}
{\bf{\large Description of Critical Points:}}
\begin{itemize}
\item At the background level, the critical points $B_{1}$, $B_{2}$, and $B_{3\pm}$ show similar behavior respectively to that of the critical points $A_{1}$, $A_{2}$, and $A_{3\pm}$ {Same as in \bf Case-I}.

\item The scaling solutions of density parameters for the critical point $B_{5}$ are $\Omega_{de}=\frac{12}{n^{2}}$, $\Omega_{m}=1-\frac{12}{n^{2}}$ and $\Omega_{r}=0$, which indicates the non-standard phase of matter-dominated Universe. The scaling solution shows the evolution between the matter-dark energy-dominated eras of the Universe. It shows the decelerated era of the Universe. The existence condition of $B_{5}$ is $n^2 > 12$. 

\item The density parameter scaling solution at the critical point $B_{7}$ is, $\Omega_{de}=\frac{16}{n^{2}}$, $\Omega_{m}=0$, and $\Omega_{r}=1-\frac{16}{n^{2}}$. Scaling solutions represent the evolution of the Universe between radiation-dark energy phases. This indicates the non-standard radiation and decelerating phase of the Universe. It exists for $n^{2}\geq 16$.

\item The critical point $B_{9}$ shows the dark energy-dominated phase of the Universe. The value of the deceleration parameter for this critical point is $\frac{1}{8} \left(n^2-8\right)$. The condition for accelerated era is, $n^2<8 $ whereas for decelerated era, $n<-2 \sqrt{2}$ and $ n>2 \sqrt{2}$. Both the EoS parameters become $\omega_{de}=\omega_{tot}=-1+\frac{n^2}{12}$. This point shows quintessence phase for $-2 \sqrt{2}<n<0 $ and $ 0<n<2 \sqrt{2}$. The phantom phase is not visible at this critical point as it does not satisfy $\omega_{tot}<-1$ for any value of $n$.   
\end{itemize}

{\bf{\large Stability Analysis:}}

The eigenvalues of the Jacobean matrix for each critical point are given below:

\begin{itemize}
\item Eigenvalues of critical point $B_{1}$
\begin{eqnarray*}
\mu_{1} = -\frac{1}{2}, \hspace{0.2cm} \mu_{2} = -\frac{3}{2}, \hspace{0.2cm} \mu_{3} = -\frac{9}{2} , \hspace{0.2cm} \mu_{4} = \frac{3}{2}\,, \hspace{0.2cm} \mu_{5} = 0  \,.   
\end{eqnarray*}
The presence of the positive eigenvalue shows the unstable behavior.
\item Eigenvalues of critical point $B_{2}$
\begin{eqnarray*}
\mu_{1} = -2, \hspace{0.2cm} \mu_{2} = -3, \hspace{0.2cm} \mu_{3} = -3 , \hspace{0.2cm} \mu_{4} = -3, \hspace{0.2cm} \mu_{5} = 0\,.  
\end{eqnarray*}

The eigenvalues of this critical point are zero, and the negative real part is known as non-hyperbolic. The stability of this point cannot be explained by linear stability theory and needs to be obtained through center manifold theory (CMT) [\ref{Sec-app}]. With this approach, the point shows stable behavior.

\item Eigenvalues of critical points $B_{3+}$ and $B_{4+}$
\begin{eqnarray*}
\mu_{1} = 3, \hspace{0.2cm} \mu_{2} = 1, \hspace{0.2cm} \mu_{3} = -6-\sqrt{6} \alpha , \hspace{0.2cm} \mu_{4} = \frac{1}{4} \left(\sqrt{6} n+12\right), \hspace{0.2cm} \mu_{5} = -\sqrt{6} \,.  
\end{eqnarray*}
These critical points indicate saddle behavior.

\item Eigenvalues of critical point $B_{3-}$ and $B_{4-}$:
\begin{eqnarray*}
\mu_{1} = 3, \hspace{0.2cm} \mu_{2} = 1, \hspace{0.2cm} \mu_{3} = -6 +\sqrt{6} \alpha , \hspace{0.2cm} \mu_{4} = 3-\frac{1}{2} \sqrt{\frac{3}{2}} n, \hspace{0.2cm} \mu_{5} = \sqrt{6} \,. 
\end{eqnarray*}
The critical points show saddle behavior for the condition $\alpha <\sqrt{6}$ and $ n>2 \sqrt{6}$, else unstable behavior.
    
\item Eigenvalues of critical point $B_{5}$ and $B_{6}$
\begin{eqnarray*}
&\mu_{1} = -\frac{1}{2}, \hspace{0.2cm} \mu_{2} =-3+ \frac{6 \alpha }{n}, \hspace{0.2cm} \mu_{3} =\frac{Root[\mathcal{Q}\&,1]}{4 n^4}, \nonumber \\&\mu_{4} =\frac{Root[\mathcal{Q}\&,2]}{4 n^4}, \hspace{0.2cm} \mu_{5} =\frac{Root[\mathcal{Q}\&,3]}{4 n^4} \,, \nonumber \\ 
&\mathcal{Q}=\#1^3+\#1^2 \left(6 n^4-24 n^3\right)+\#1 \left(72 n^8-216 n^7-864 n^6\right)\nonumber \\&-864 n^{11}+10368 n^9
\end{eqnarray*} 

We have obtained the above output in Mathematica, where $Root[\mathcal{Q}\&,1]$, $Root[\mathcal{Q}\&,2]$, and $Root[\mathcal{Q}\&,3]$ represents the first, second, and third roots of the function $\mathcal{Q}$ respectively. The $\#1$  represented is an anonymous function with coefficients that depend on $n$. So, these critical point shows unstable behavior.

\item Eigenvalues of critical point $B_{7}$ and $B_{8}$
\begin{eqnarray*}
&\mu_{1} = 1, \hspace{0.2cm} \mu_{2} =-4+ \frac{8 \alpha }{n}, \hspace{0.2cm} \mu_{3} =\frac{Root[\mathcal{P}\&,1]}{12 n^4}, \nonumber \\& \mu_{4} =\frac{Root[\mathcal{P}\&,2]}{12 n^4}, \hspace{0.2cm} \mu_{5} =\frac{Root[\mathcal{P}\&,3]}{12 n^4} \,, \nonumber \\ &
\mathcal{P}=\#1^3+\#1^2 \left(12 n^4-96 n^3\right)+\#1 \left(576 n^8-1728 n^7-9216 n^6\right)\nonumber \\&-27648 n^{11}+442368 n^9.
\end{eqnarray*}
According to the behavior of the eigenvalues, these critical point indicates unstable behavior.

\item Eigenvalues of critical point $B_{9}$ and $B_{10}$
\begin{eqnarray*}
\mu_{1} = -2+\frac{n^2}{8}, \hspace{0.4cm} \mu_{2} = -3+\frac{n^2}{4} , \hspace{0.4cm} \mu_{3} =-\frac{1}{4} n (n-2 \alpha )\,,\nonumber \\
\mu_{4} = \frac{1}{16} \left(n (n+4)-\sqrt{((n-12) n-72) ((n-4) n-8)}-24\right) \,, \nonumber \\
\mu_{5} = \frac{1}{16} \left(n (n+4)+\sqrt{((n-12) n-72) ((n-4) n-8)}-24\right) \,.
\end{eqnarray*}
The critical points $B_{9}$ and $B_{10}$ show stable  behavior for the conditions $2 \left(1-\sqrt{3}\right)\leq n<0$ and $\alpha >\frac{n}{2}$. The stability of this point depends on the parameters $\alpha$ and $n$. The stability region plot is given in FIG.-\ref{Figb9b10M2}, where the green/shaded region shows the stable phase.   
 \begin{figure}[H]
 \centering
 \includegraphics[width=100mm]{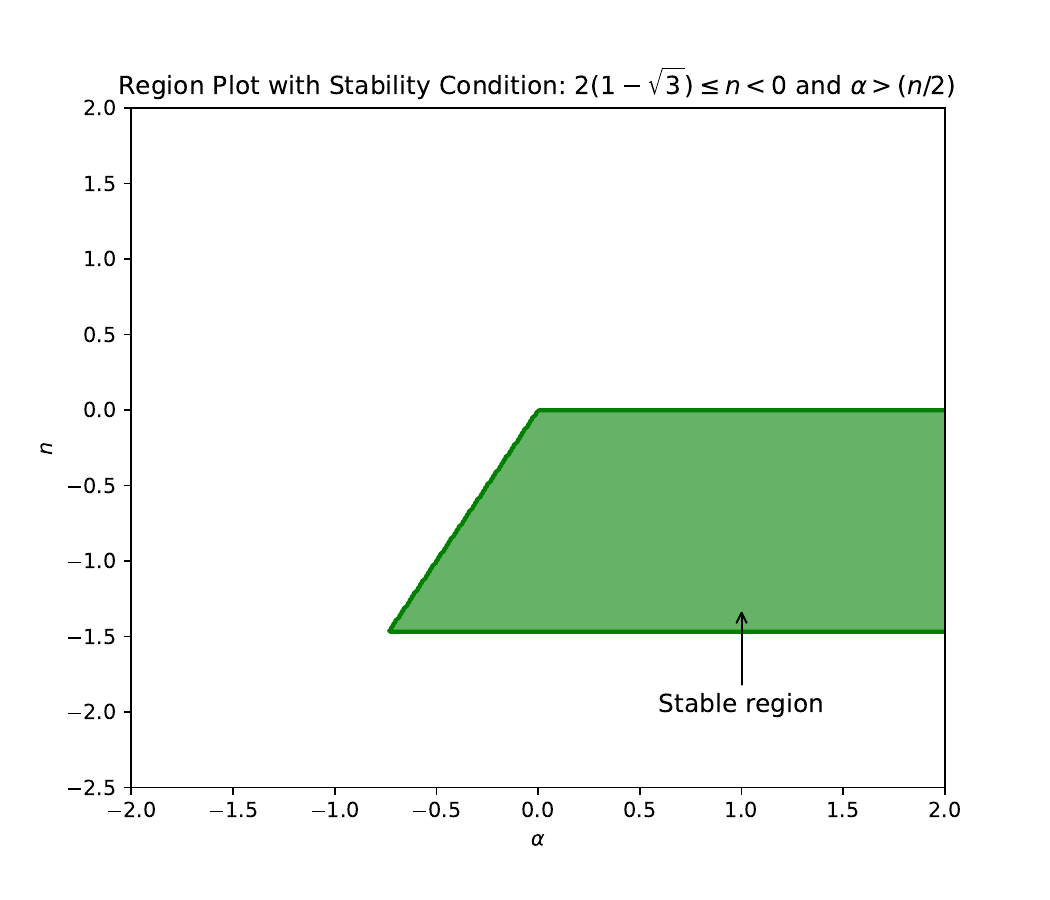}
 \caption{Stability region of the critical points $B_{9}$ and $B_{10}$  between the model parameter $\alpha$ and $n$.} \label{Figb9b10M2}
 \end{figure}
\end{itemize}

{\bf{\large Numerical Solutions:}}\\

In this case, three attractor points ($B_{2}$, $B_{9}$, $B_{10}$) are obtained which describe the accelerating phase of the Universe. \\

\begin{figure}
     \centering
     \begin{subfigure}[b]{0.49\textwidth}
         \centering
         \includegraphics[width=\linewidth]{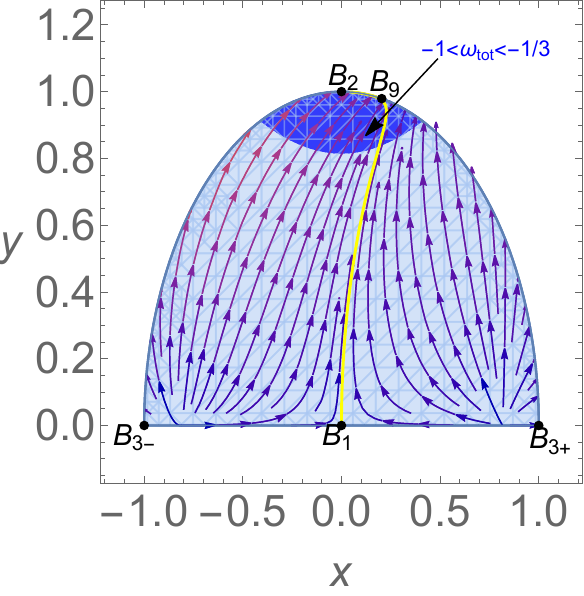}
         \caption{Phase space with $\alpha=-0.6$, $n=-1$}
         \label{fig:case2_phaseportrait}
     \end{subfigure}
     \hfill
     \begin{subfigure}[b]{0.49\textwidth}
         \centering
         \includegraphics[width=\linewidth]{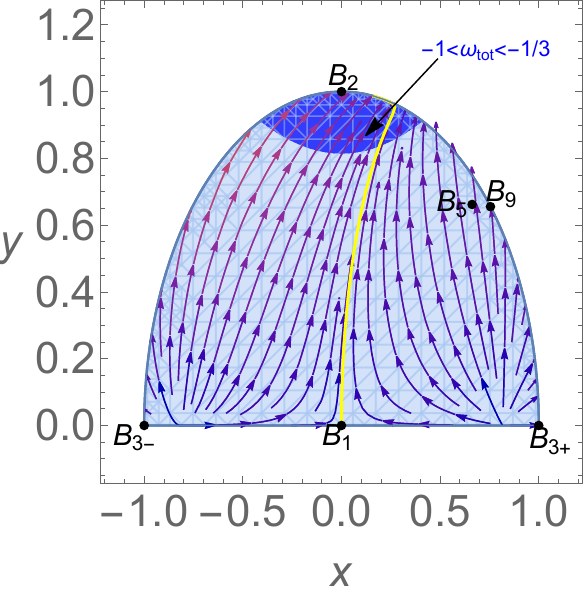}
         \caption{Phase space with $\alpha=-0.6$, $n=-3.7$}
         \label{fig:case2_phaseB5}
     \end{subfigure}
     \hfill
     \begin{subfigure}[b]{0.49\textwidth}
         \centering
         \includegraphics[width=\linewidth]{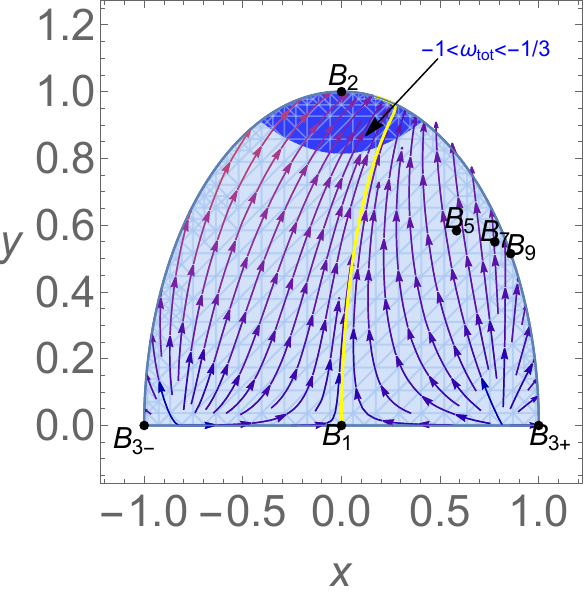}
         \caption{Phase space with $\alpha=-0.6$, $n=-4.2$}
         \label{fig:case2_phaseB7}
     \end{subfigure}
\caption{CaseII: 2D phase space portrait for the autonomous system (\ref{autonomous-system1}-\ref{autonomous-system6}). The blue/shaded region represents the portion of the phase space where the universe undergoes accelerated expansion $(-1<\omega_{tot}<-\frac{1}{3})$. The initial conditions are: $x = 10^{-5}$, $y = 9 \times 10^{-13}$, $u=10^{-5}$, $\rho=\sqrt{0.999661}$, $\lambda=0.8$.} 
 \label{Fig4phase_portrait}
\end{figure}

In FIG. \ref{Fig4phase_portrait}, we examine the phase space portrait for different values of the model parameter $n$. Based on the value of $n$, the phase space has been divided into three regions. The variation in $n$ emphasizes the presence of the critical points $B_5-B_8$ and an accelerating region (blue/shaded) associated with the critical points $B_9$ and $B_{10}$.
\\

In FIG. \ref{fig:case2_phaseportrait}, we illustrate the phase portrait for $n = -1$. At this value, the critical points $B_5$–$B_8$ do not exist because the critical points $B_5$–$B_6$ and $B_7$–$B_8$ only emerge when $n^2 > 12$ and $n^2 > 16$ respectively. These conditions are consistent with the physical viability of the density parameters $(0 < \Omega_r < 1)$ and $(0 < \Omega_m < 1)$. In the accelerating region associated with the critical points $B_9$ and $B_{10}$ where $n^2 < 8$, the critical points $B_5$–$B_6$ and $B_7$–$B_8$ do not appear. For $n = -1$, the critical point $B_9$ lies within the accelerating region (blue/shaded). Specifically, the critical point $B_9$ corresponds to the accelerating phase when $n^2 < 8$. Thus, we have chosen $n = -1$ since it satisfies the stability conditions for a critical point $B_9$.\\

All the trajectories in the phase space are heteroclinic orbits that begins at $B_{3\pm}$ and ends at $B_9$. There are two heteroclinic orbits, $B_{3\pm} \to B_1 \to B_9$. These orbits can serve as physical models for the transition from dark matter to dark energy, effectively characterizing the late-time evolution of the Universe. The total EoS, $\omega_{tot} = -1 + \frac{n^2}{12}$. However at early times the model always predicts stiff fluid domination, represented by $B_{3\pm}$ and it does not favour from the phenomenological perspective.\\

For $n^2 > 8$, point $B_9$ lies outside the acceleration region (blue/shaded) and do not represent an inflationary solution. At $n=-1$, the critical point $B_9$ exhibits stable node behavior and indicates the late-time cosmic acceleration of the Universe. The heteroclinic orbit solution (yellow line) is obtained from the numerical solution of the autonomous system (\ref{autonomous-system1}-\ref{autonomous-system6}) with the initial conditions $x = 10^{-5}$, $y = 9 \times 10^{-13}$, $u = 10^{-5}$, $\rho = \sqrt{0.999661}$, and $\lambda = 0.8$.\\

In FIG. \ref{fig:case2_phaseB5}, we observe six critical points in the phase space under the condition $n^2 > 12$. For $n = -3.7$, the critical points $B_7$ and $B_8$ do not appear, and the critical point $B_9$ lies outside the accelerating region (blue/shaded). At $n=-3.7$, the critical points $B_5$, $B_6$ and $B_9$ exhibit saddle-like (unstable) behavior, indicating a decelerating phase of the Universe. Point $B_9$ consistently falls outside the acceleration region (blue/shaded) and thus never represents an inflationary solution. There are two heteroclinic orbits: $B_{3\pm} \to B_1 \to B_2$.
\\

In FIG. \ref{fig:case2_phaseB7}, we identify seven critical points in the phase space for the condition $n^2 > 16$. For $n = -4.2$, the critical point $B_9$ is positioned outside the accelerating region (blue/shaded). At this value, the critical points $B_5$, $B_7$ and $B_9$ exhibit saddle-like (unstable) behavior, signaling the decelerating phase of the Universe. Importantly, point $B_9$ lies outside the acceleration region (blue/shaded) and, therefore, cannot represent an inflationary solution. There are two heteroclinic orbits: $B_{3\pm} \to B_1 \to B_2$.
 \\

\begin{figure}
     \centering
     \begin{subfigure}[b]{0.4\textwidth}
         \centering
         \includegraphics[width=\linewidth]{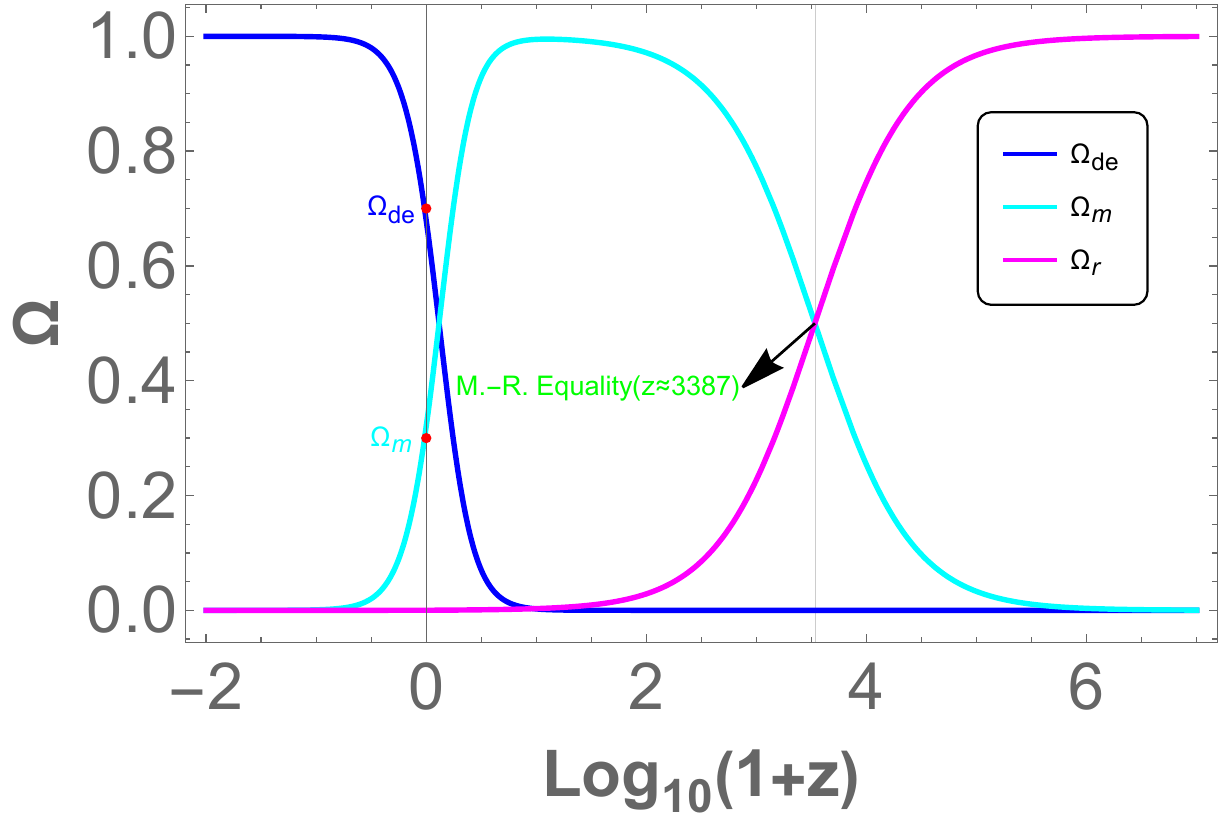}
         \caption{Evolution of density parameters.}
         \label{fig:case2_density}
     \end{subfigure}
     \hfill
     \begin{subfigure}[b]{0.4\textwidth}
         \centering
         \includegraphics[width=\linewidth]{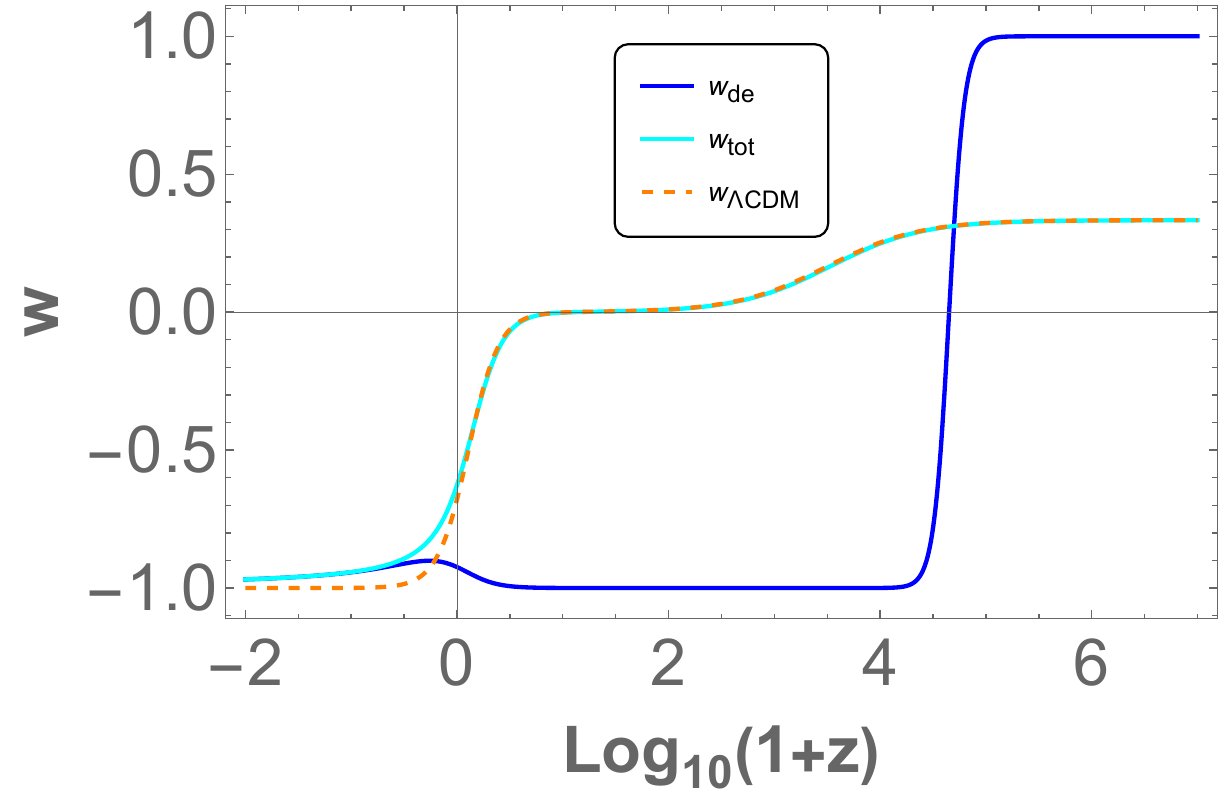}
         \caption{Evolution of EoS parameters.}
         \label{fig:case2_statepara}
     \end{subfigure}
\caption{In this figure, we set $\alpha=-0.6$ and $n=-1$ with the initial conditions are the same as in FIG.- \ref{Fig4phase_portrait}.} 
\label{Fig3}
\end{figure}

\begin{figure}
     \centering
     \begin{subfigure}[b]{0.4\textwidth}
         \centering
         \includegraphics[width=\linewidth]{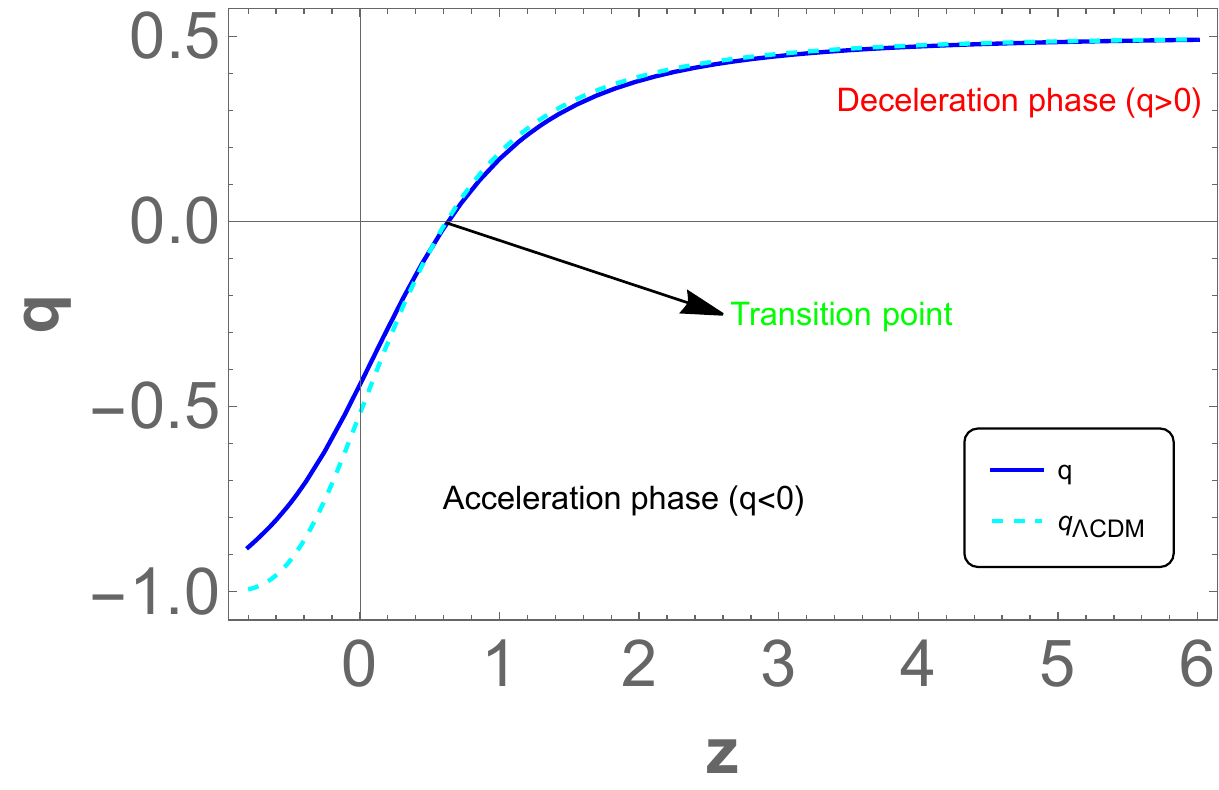}
         \caption{Evolution of deceleration parameter $q$ in redshift.}
         \label{fig:case2_qz}
     \end{subfigure}
     \hfill
     \begin{subfigure}[b]{0.4\textwidth}
         \centering
         \includegraphics[width=\linewidth]{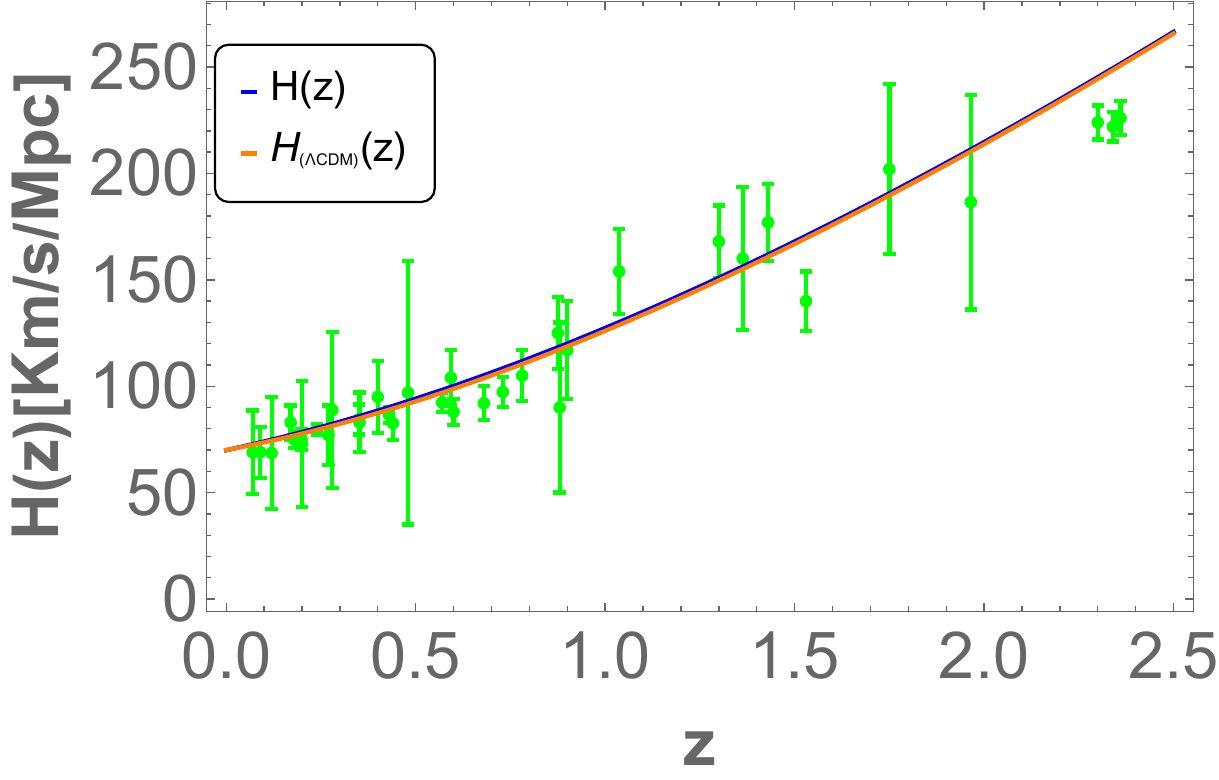}
         \caption{Evolution of the Hubble rate $H(z)$ in redshift .}
         \label{fig:case2_Hz}
     \end{subfigure}
\caption{In this figure, we set $\alpha=-0.6$ and $n=-1$ with the initial conditions are the same as in FIG.- \ref{Fig4phase_portrait}.} 
\label{Fig4}
\end{figure}

\begin{figure}[H]
 \centering
 \includegraphics[width=100mm]{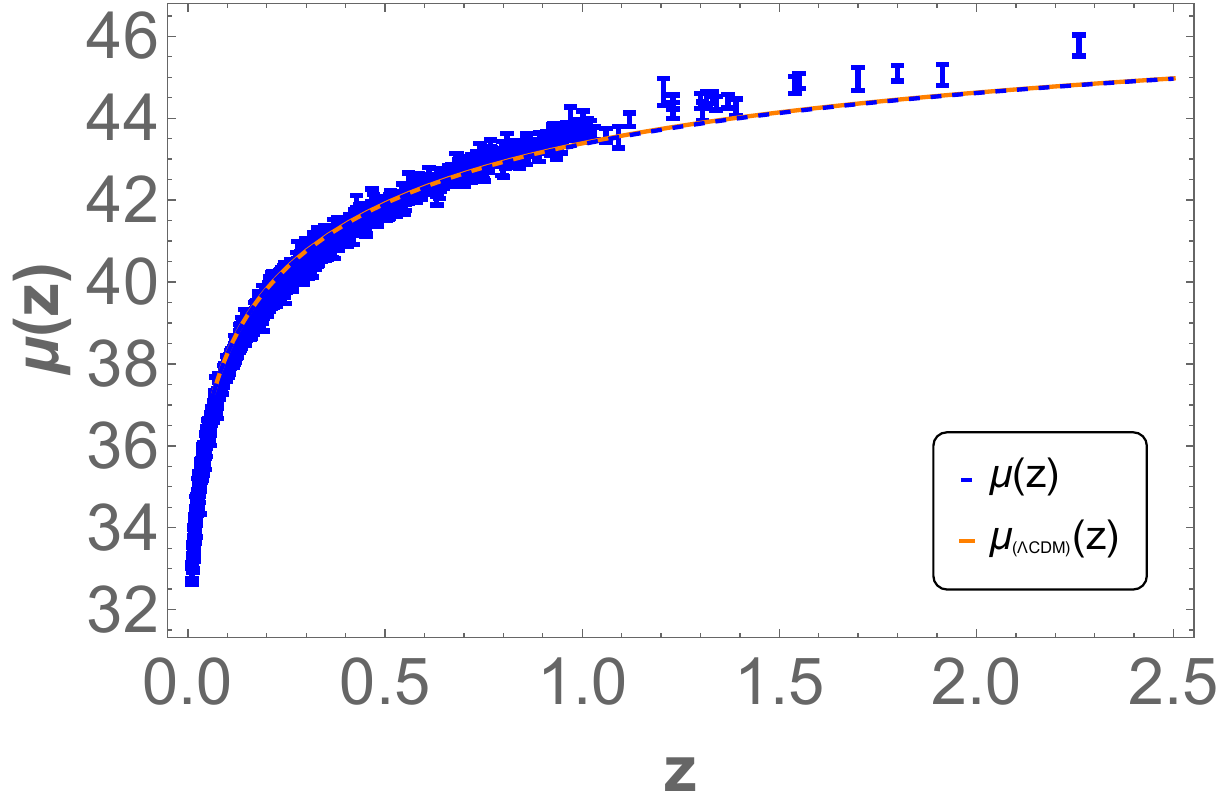} 
 \caption{We show the evolution of the distance modulus function $\mu(z)$ ( dashed blue) and the $\Lambda$CDM model distance modulus function $\mu_{\Lambda CDM}(z)$ along with the 1048 Supernovae Ia (SNe Ia)  data points \cite{Scolnic_2018}. In this figure, we set $\alpha=-0.6$ and $n=-1$ with the initial conditions are the same as in FIG.- \ref{Fig4phase_portrait}.} \label{Fig4pan}
 \end{figure}

 For these solutions to be cosmologically viable, it is essential that the potential be sufficiently flat ($n^2 < 8$) and the initial conditions are fine-tuned to ensure the prolonged domination of dark energy. The solution should follow the sequence: $B_{3\pm} \to B_1 \to B_9$ (see FIG.~\ref{fig:case2_phaseportrait}). Further at early times, the only admissible solutions correspond to non-physical stiff fluid Universe fails to describe accurately. The EoS parameters, energy densities, deceleration parameters, Hubble rate, and the modulus function for a solution that shadows the heteroclinic sequence $B_{3\pm} \to B_1 \to B_9$ are shown in FIG.-\ref{Fig3}, FIG.-\ref{Fig4}, FIG.-\ref{Fig4pan}.
\\

FIG.-\ref{fig:case2_density} shows the evolutionary behavior of radiation, dark energy, and dark matter. The radiation appears first in the early cosmos then the dark matter takes over for a short while and eventually the cosmological constant. The present value of density parameters, $\Omega_{m}\approx 0.33$, $\Omega_{de}\approx 0.67$ and the matter-radiation equality at $z\approx 3387$. In FIG.-\ref{fig:case2_statepara}, the evolution of EoS parameters are shown. We observe that $\omega_{tot}$ (cyan) begins at $\frac{1}{3}$ for radiation, decreases to 0 during the matter-dominated period and ultimately reaches approximately $-1$. The EoS parameter $\omega_{\Lambda CDM}$ and the dark energy dominated EoS parameter $\omega_{de}$ (blue) approach about to $-1$ at late-time. The current value of the EoS parameter for the dark energy sector, $\omega_{de}(z=0)=-0.92$, is consistent with the current Planck collaboration [$\omega_{de}(z=0)= -1.028 \pm 0.032$ \cite{Aghanim:2018eyx}]. The deceleration parameter [FIG.-\ref{fig:case2_qz}] shows the transition at $z\approx 0.62$ from decelerating to accelerating aligning with the recent findings \cite{PhysRevD.90.044016a}. The present value, $q(z=0) \approx -0.45$, agrees with the visualized cosmic identification \cite{PhysRevResearch.2.013028}. The visualization of the Hubble rate evolution, Hubble rate $H_{\Lambda CDM}(z)$  and the Hubble data points \cite{Moresco_2022_25} are shown in FIG.-\ref{fig:case2_Hz}. We have used $H_{0}=70$ Km/(Mpc sec) as the current Hubble parameter \cite{Aghanim:2018eyx}. The outcomes of the standard $\Lambda$CDM model are extremely similar to the obtained result. The evolution of the modulus function $\mu(z)$ is presented in FIG.-\ref{Fig4pan} along with the $\Lambda$CDM model modulus function $\mu_{\Lambda CDM}$ and 1048 pantheon data points. The mathematical formalism of the Hubble and Pantheon data sets is presented in section--\ref{dataset}.
\\
 
\subsection{{\large Case-III:} $V(\phi)=V_{0}\sinh^{-\gamma}(\beta \phi)$ \cite{SAHNI_2000, Ure_a_L_pez_2016}} \label{case3sinh}
In this case, the roots of the function $f(\lambda)$ are, 
\begin{eqnarray}\label{sinnhpotentialroot}
f(\lambda)&=&\frac{\lambda^{2}}{\gamma}-\gamma \beta^{2}\,, \hspace{0.5cm} \beta_{1}=\frac{1}{\gamma}\,, \hspace{0.5cm} \beta_{2}=0 \,, \hspace{0.2cm} \beta_{3}=-\gamma \beta^{2}\,. \nonumber \\&&
\lambda^{+}= \gamma\beta \,,\hspace{1cm}\lambda^{-}=-\gamma \beta,
\end{eqnarray}

In TABLE-\ref{TABLE-III}, the eigenvalues of the critical points and their corresponding existence conditions are given. The values of the cosmological parameters are listed in  TABLE- \ref{TABLE-IV}. The eigenvalues and background cosmological parameters for the critical points $B_{1}$ and $B_{2}$ are obtained to be the same as that of the power-law potential function ($V(\phi)=V_{0}\phi^{n}$) {\bf [Case-II]]}. Since, the model parameter does not affect the critical points $B_{1}$ and $B_{2}$, so only the other critical points will be investigated.\\

{\bf{\large Description of Critical Points:}}

\begin{itemize}
\item  For the critical points $B_{3\pm}$ and $B_{4\pm}$,  $\omega_{de}=\omega_{tot}=1$, $\Omega_{de}=1$ and so these points described the stiff-matter. The positive value of $q$ indicates the deceleration phase. 

\item The density parameters scaling solutions of the critical points $B_{5}$ and $B_{6}$ are $\Omega_{de}=\frac{3}{\gamma^2 \beta^2}$ and $\Omega_{m}=1-\frac{3}{\gamma^2 \beta^2}$. Also, we have  $\omega_{de}=\omega_{tot}=0$. The scaling solution shows the evolution between matter-dark energy-dominated phases of the Universe. These critical points show a deceleration phase, indicated by the positive value of the deceleration parameter. The points $B_5$ and $B_6$ exist if they satisfy the condition $\gamma^2 \beta^2 > 3$.

\item The critical points $B_{7}$ and $B_{8}$ have density parameter scaling solutions of $\Omega_{de}=\frac{4}{\gamma^2 \beta^2}$ and $\Omega_{r}=1-\frac{4}{\gamma^2 \beta^2}$. The critical points reveal non-standard radiation and the decelerating phase of the Universe. Scaling solutions illustrate the transition between radiation-dark energy-dominated phases of the Universe. The existence of the points $B_7$ and $B_8$ is contingent upon the condition $\gamma^2 \beta^2 > 4$ being satisfied.

\item The critical points $B_{9}$ and $B_{10}$ reveal the dark energy dominated era of the Universe. The solution of the dark energy and total EoS parameters is the same, i.e. $\omega_{de}=\omega_{tot}=-1+\frac{\beta ^2 \gamma ^2}{3}$ and the solution of the deceleration parameter is $q=-1+\frac{\beta ^2 \gamma ^2}{2}$. The points show the accelerated phase for the conditions $\left(\gamma <0\land -\sqrt{2} \sqrt{\frac{1}{\gamma ^2}}<\beta <\sqrt{2} \sqrt{\frac{1}{\gamma ^2}}\right)$, $\gamma=0$, and $\bigg(\gamma >0\land -\sqrt{2} \sqrt{\frac{1}{\gamma ^2}}<\beta <\sqrt{2} \sqrt{\frac{1}{\gamma ^2}}\bigg)$;  decelerated phase for the conditions $\bigg(\gamma <0\land \left(\beta <-\sqrt{2} \sqrt{\frac{1}{\gamma ^2}}\lor \beta >\sqrt{2} \sqrt{\frac{1}{\gamma ^2}}\right)\bigg)$ and $\bigg(\gamma >0\land \bigg(\beta <-\sqrt{2} \sqrt{\frac{1}{\gamma ^2}}\lor \beta >\sqrt{2} \sqrt{\frac{1}{\gamma ^2}}\bigg)\bigg)$. For the condition on the model parameter $\bigg(\beta <0\land \bigg(-\sqrt{2} \sqrt{\frac{1}{\beta ^2}}<\gamma <0\lor 0<\gamma <\sqrt{2} \sqrt{\frac{1}{\beta ^2}}\bigg)\bigg)$ and $\bigg(\beta >0\land \bigg(-\sqrt{2} \sqrt{\frac{1}{\beta ^2}}<\gamma <0\lor 0<\gamma <\sqrt{2} \sqrt{\frac{1}{\beta ^2}}\bigg)\bigg)$ points $B_{9}$ and $B_{10}$ imply the quintessence phase of the Universe. In Fig.~\ref{fig:case3_q_region}, we present the region plot of the model parameters $\beta$ and $\gamma$, showing the deceleration parameter for both the acceleration phase ($q < 0$) and the deceleration phase ($q > 0$).  In Fig.~\ref{fig:case3_omega_region}, we show the region plot of the model parameters $\beta$ and $\gamma$ for the quintessence phase ($-1 < \omega_{tot} < -\frac{1}{3}$).

\begin{figure}
     \centering
     \begin{subfigure}[b]{0.4\textwidth}
         \centering
         \includegraphics[width=\linewidth]{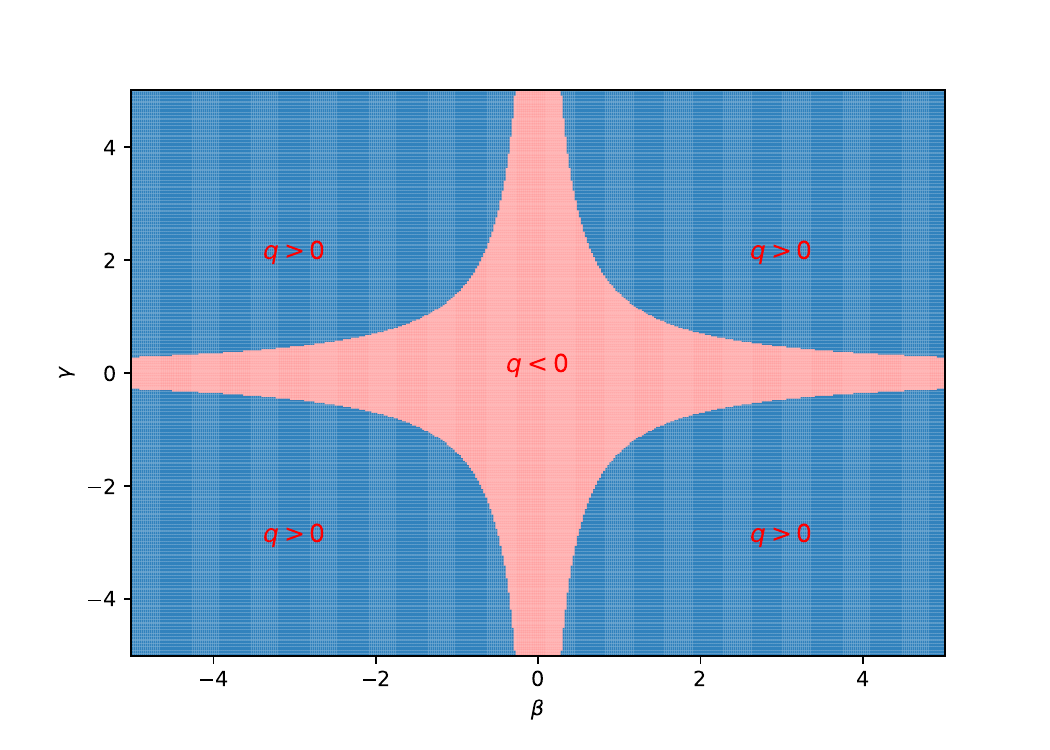}
         \caption{Region plot between the model parameters $\gamma$ and $\beta$ for the deceleration parameter}
         \label{fig:case3_q_region}
     \end{subfigure}
     \hfill
     \begin{subfigure}[b]{0.4\textwidth}
         \centering
         \includegraphics[width=\linewidth]{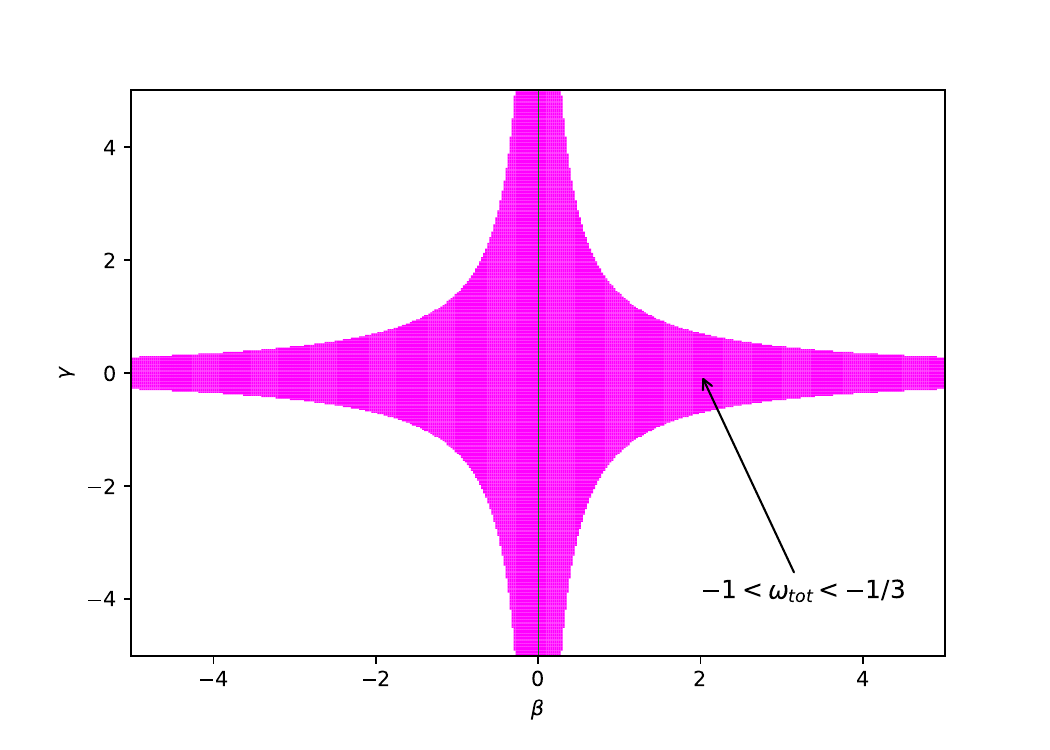}
         \caption{Region plot between the model parameters $\gamma$ and $\beta$ for the quintessence phase.  }
         \label{fig:case3_omega_region}
     \end{subfigure}
\caption{Region plot between the model parameters $\gamma$ and $\beta$ for the critical points $B_{9}$ and $B_{10}$ for Case-III.} 
\label{Figb9b10}
\end{figure}
\end{itemize}

{\bf{\large Stability Analysis:}}

\begin{itemize}
\item Eigenvalues of the critical point $B_{3\pm}$ and $B_{4\pm}$
\begin{eqnarray*}
\mu_{1} = 3, \hspace{0.2cm} \mu_{2} = 1, \hspace{0.2cm} \mu_{3} = -6\pm\sqrt{6} \alpha , \hspace{0.2cm} \mu_{4} = \pm2 \sqrt{6} \beta, \hspace{0.2cm} \mu_{5} = 3\pm\sqrt{\frac{3}{2}} \beta  \gamma \,.  
\end{eqnarray*}
Sign($-$) in $\mu_{3}$, $\mu_{4}$, and $\mu_{5}$ indicate the eigenvalues of the critical points $B_{3+}$ and $B_{4-}$, whereas Sign ($+$) in $\mu_{3}$, $\mu_{4}$, and $\mu_{5}$ imply the eigenvalues of the critical points $B_{3-}$ and $B_{4+}$. The critical points $B_{3+}$ and $B_{4-}$ indicate the saddle behavior for the conditions $\alpha >-\sqrt{6}\land \gamma >0\land \beta >\frac{\sqrt{6}}{\gamma } $, nonetheless it is showing unstable node behaviour. The critical points $B_{3-}$ and $B_{4+}$ show the saddle behavior for the conditions $\alpha <\sqrt{6}\land \gamma >0\land \beta <-\frac{\sqrt{6}}{\gamma }$; else it is defined as an unstable node. 
    
\item Eigenvalues of the critical point $B_{5}$ and $B_{6}$
\begin{eqnarray*}
\mu_{1} = -\frac{6}{\gamma }, \hspace{0.5cm} \mu_{2} =-\frac{1}{2}, \hspace{0.5cm} \mu_{3} =-3\pm\frac{3 \alpha }{\beta  \gamma }, \nonumber \\  \mu_{4} =-\frac{3}{4}-\frac{3 \sqrt{24 \beta ^6 \gamma ^6-7 \beta ^8 \gamma ^8}}{4 \beta ^4 \gamma ^4}, \hspace{0.5cm}
\mu_{5} =-\frac{3}{4}+\frac{3 \sqrt{24 \beta ^6 \gamma ^6-7 \beta ^8 \gamma ^8}}{4 \beta ^4 \gamma ^4} \,.
\end{eqnarray*}
Sign($-$) and ($+$) in the $\mu_{3}$ imply the eigenvalues of the critical point $B_{5}$ and $B_{6}$ respectively. Both $B_{5}$ and $B_{6}$ are showing stable behavior for the conditions 
$\bigg(-2 \sqrt{\frac{6}{7}}<\alpha \leq -\sqrt{3}\land \gamma >0\land \bigg(-2 \sqrt{\frac{6}{7}} \sqrt{\frac{1}{\gamma ^2}}\leq \beta <\frac{\alpha }{\gamma }\lor -\frac{\alpha }{\gamma }<\beta \leq 2 \sqrt{\frac{6}{7}} \sqrt{\frac{1}{\gamma ^2}}\bigg)\bigg)\lor  \bigg(-\sqrt{3}<\alpha <\sqrt{3}\land \gamma >0\land \bigg(-2 \sqrt{\frac{6}{7}} \sqrt{\frac{1}{\gamma ^2}}\leq \beta <-\sqrt{3} \sqrt{\frac{1}{\gamma ^2}}\lor \sqrt{3} \sqrt{\frac{1}{\gamma ^2}}<\beta \leq 2 \sqrt{\frac{6}{7}} \sqrt{\frac{1}{\gamma ^2}}\bigg)\bigg)\lor \bigg(\sqrt{3}\leq \alpha <2 \sqrt{\frac{6}{7}}\land \gamma >0\land \bigg(-2 \sqrt{\frac{6}{7}} \sqrt{\frac{1}{\gamma ^2}}\leq \beta <-\frac{\alpha }{\gamma }\lor \frac{\alpha }{\gamma }<\beta \leq 2 \sqrt{\frac{6}{7}} \sqrt{\frac{1}{\gamma ^2}}\bigg)\bigg)$. It is saddle or unstable if the critical points do not satisfy the above conditions. 
 
\item Eigenvalues of the critical point $B_{7}$ and $B_{8}$
\begin{eqnarray*}
\mu_{1} = -\frac{8}{\gamma }, \hspace{0.5cm} \mu_{2} =1, \hspace{0.5cm} \mu_{3} =-4\pm\frac{4 \alpha }{\beta  \gamma }, \nonumber \\  \mu_{4} =-\frac{1}{2}-\frac{\sqrt{64 \beta ^6 \gamma ^6-15 \beta ^8 \gamma ^8}}{2 \beta ^4 \gamma ^4}, \hspace{0.5cm}
\mu_{5} =-\frac{1}{2}+\frac{\sqrt{64 \beta ^6 \gamma ^6-15 \beta ^8 \gamma ^8}}{2 \beta ^4 \gamma ^4} \,.
\end{eqnarray*}
Indices ($-$) and ($+$) in $\mu_{3}$ denote the eigenvalues of the critical points $B_{5}$ and $B_{6}$ respectively. The points $B_{5}$ and $B_{6}$ imply the saddle behavior for the condition $\bigg(-\frac{8}{\sqrt{15}}<\alpha \leq -2\land \bigg(\bigg(\beta <0\land \frac{\alpha }{\beta }<\gamma \leq \frac{8 \sqrt{\frac{1}{\beta ^2}}}{\sqrt{15}}\bigg)\lor \bigg(\beta >0\land -\frac{\alpha }{\beta }<\gamma \leq \frac{8 \sqrt{\frac{1}{\beta ^2}}}{\sqrt{15}}\bigg)\bigg)\bigg)\lor \bigg(-2<\alpha \leq 2\land \bigg(\bigg(\beta <0\land -\frac{2}{\beta }<\gamma \leq \frac{8 \sqrt{\frac{1}{\beta ^2}}}{\sqrt{15}}\bigg)\lor \bigg(\beta >0\land \frac{2}{\beta }<\gamma \leq \frac{8 \sqrt{\frac{1}{\beta ^2}}}{\sqrt{15}}\bigg)\bigg)\bigg)\lor \bigg(2<\alpha <\frac{8}{\sqrt{15}}\land \bigg(\bigg(\beta <0\land -\frac{\alpha }{\beta }<\gamma \leq \frac{8 \sqrt{\frac{1}{\beta ^2}}}{\sqrt{15}}\bigg)\lor \bigg(\beta >0\land \frac{\alpha }{\beta }<\gamma \leq \frac{8 \sqrt{\frac{1}{\beta ^2}}}{\sqrt{15}}\bigg)\bigg)\bigg)$. Except above conditions the points shows unstable node behavior.

\item Eigenvalues of the critical points $B_{9}$ and $B_{10}$\\
Both the critical points have same eigenvalues for $\mu_{1}$, $\mu_{2}$, $\mu_{3}$, and $\mu_{4}$ and only different is in the eigenvalue $\mu_{5}$. 
\begin{eqnarray*}
\mu_{1} = -2 \beta ^2 \gamma, \hspace{0.2cm} \mu_{2} =-3+\frac{\beta ^2 \gamma ^2}{2}, \hspace{0.2cm} \mu_{3} =-2+\frac{\beta ^2 \gamma ^2}{2} \,, \hspace{0.2cm} \mu_{4} =-3+\beta ^2 \gamma ^2, \\
\mu_{5} =-\beta  \gamma  (\alpha +\beta  \gamma ) \hspace{0.2cm} (B_{9}) \,, \hspace{0.2cm} \mu_{5} =\beta  \gamma  (\alpha -\beta  \gamma ) \hspace{0.2cm} (B_{10})\,.
\end{eqnarray*}   
The points show stable node behavior for the conditions 
$\bigg(-\sqrt{3}<\alpha \leq 0\land \bigg(\bigg(\beta <0\land \frac{\alpha }{\beta }<\gamma <\sqrt{3} \sqrt{\frac{1}{\beta ^2}}\bigg)\lor \left(\beta >0\land -\frac{\alpha }{\beta }<\gamma <\sqrt{3} \sqrt{\frac{1}{\beta ^2}}\right)\bigg)\bigg)\lor \bigg(0<\alpha <\sqrt{3}\land \bigg(\bigg(\beta <0\land -\frac{\alpha }{\beta }<\gamma <\sqrt{3} \sqrt{\frac{1}{\beta ^2}}\bigg)\lor \bigg(\beta >0\land \frac{\alpha }{\beta }<\gamma <\sqrt{3} \sqrt{\frac{1}{\beta ^2}}\bigg)\bigg)\bigg)$. Points will demonstrate unstable node and saddle behavior if they fail to fulfill the above conditions. These points also demonstrate the dark energy dominance phase of the Universe.
\end{itemize}

{\bf{\large Numerical Solutions}:}\\

\begin{figure}
     \centering
     \begin{subfigure}[b]{0.49\textwidth}
         \centering
         \includegraphics[width=\linewidth]{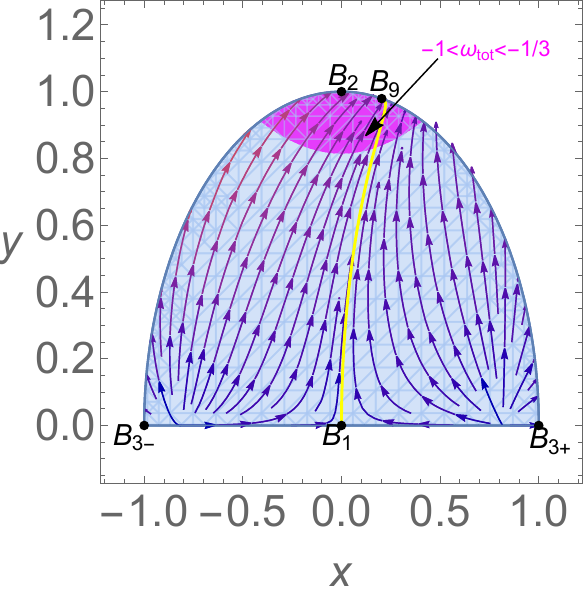}
         \caption{Phase space with $\alpha-0.1$, $\gamma=0.5$, and $\beta=1$}
         \label{fig:case3_phaseportrait}
     \end{subfigure}
     \hfill
     \begin{subfigure}[b]{0.49\textwidth}
         \centering
         \includegraphics[width=\linewidth]{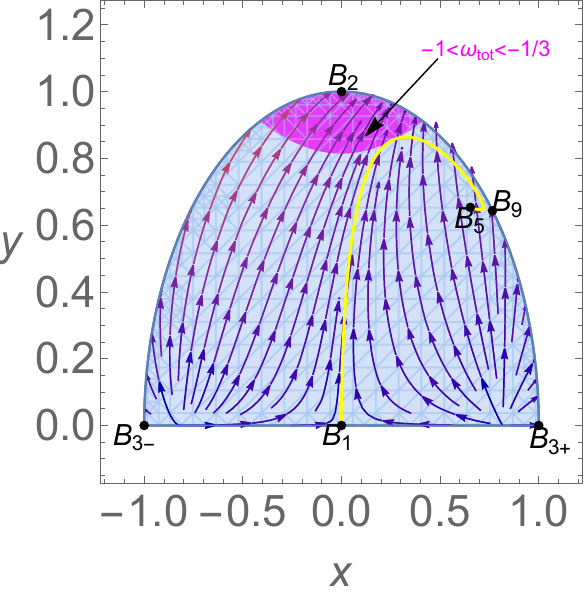}
         \caption{Phase space with $\alpha-0.1$, $\gamma=1.25$, and $\beta=1.5$}
         \label{fig:case3_phaseB5}
     \end{subfigure}
     \hfill
     \begin{subfigure}[b]{0.49\textwidth}
         \centering
         \includegraphics[width=\linewidth]{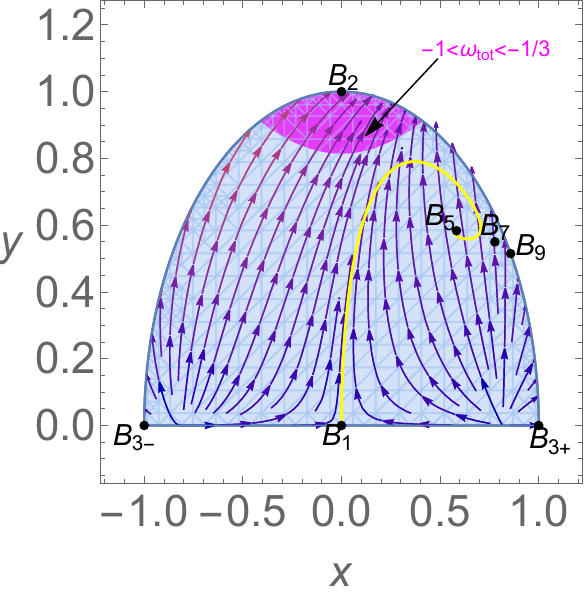}
         \caption{Phase space with $\alpha-0.1$, $\gamma=1.4$, and $\beta=1.5$}
         \label{fig:case3_phaseB7}
     \end{subfigure}
\caption{CaseIII: 2D phase space portrait plot for the autonomous system (\ref{autonomous-system1}-\ref{autonomous-system6}). The magenta/shaded region represents the portion of the phase space where the universe undergoes accelerated expansion $(-1<\omega_{tot}<-\frac{1}{3})$. The initial conditions are: $x = 10^{-5}$, $y = 9 \times 10^{-13} $, $u=10^{-5}$, $\rho=\sqrt{0.999661}$, $\lambda=0.8$.} 
\label{Fig6phase_portrait}
\end{figure}

In FIG.~\ref{Fig6phase_portrait}, the phase space portrait for varying values of the model parameters $\gamma$ and $\beta$ has been presented. The phase space structure is divided into three distinct regions, determined by the specific ranges of $\gamma$ and $\beta$. These regions represent different dynamical behaviors of the system. The variation in the values of $\gamma$ and $\beta$ reveals the existence of an accelerating region of the critical points $B_9$ and $B_{10}$, indicated by the magenta (shaded) area, and the existence of the critical points $B_5-B_{8}$. The phase space portrait illustrates how variations in $\gamma$ and $\beta$ govern the onset of cosmic acceleration, emphasizing the crucial role these parameters play in the overall evolution of the Universe.\\

In FIG.~\ref{fig:case3_phaseportrait}, we present the phase portrait for $\gamma = 0.5$ and $\beta = 1$. For these values the critical points $B_5$-$B_8$ do not exist since $\beta^2 \gamma^2 > 3$ and $\beta^2 \gamma^2 > 4$ are required for the appearance of the points $B_5$-$B_6$ and $B_7$-$B_8$ respectively. These conditions are consistent with the physical viability on the density parameters $(0 < \Omega_r < 1)$ and $(0 < \Omega_m < 1)$. In the accelerating region associated with the critical points $B_9$ and $B_{10}$, where $\beta^2 \gamma^2 < 2$, the critical points $B_5$-$B_6$ and $B_7$-$B_8$ do not exist. For $\gamma = 0.5$ and $\beta = 1$, the critical point lies within the accelerating region (magenta/shaded). The critical point $B_9$ corresponds to the accelerating phase when $\beta^2 \gamma^2 < 2$ also satisfying the stability conditions for $B_9$.

All the phase space trajectories are heteroclinic orbits starting from the point $B_{3\pm}$ and ending at $B_9$. Two such heteroclinic orbits $B_{3\pm} \to B_1 \to B_9$ provide a physical model for the transition from dark matter to dark energy. The total EoS parameter is given by $\omega_{tot} = -1 + \frac{\beta^2 \gamma^2}{3}$. However, at early times the model predicts stiff fluid domination represented by the points $B_{3\pm}$, which is not favored from a phenomenological standpoint. If $\beta^2 \gamma^2 > 2$, the point $B_9$ would lie outside the acceleration region (magenta/shaded) and would not represent an inflationary solution. 

For $\gamma = 0.5$ and $\beta = 1$, the critical point $B_9$ behaves as a stable node that shows the late-time cosmic acceleration of the Universe. The heteroclinic orbit solution (yellow line) is obtained from the numerical solution of the autonomous system (\ref{autonomous-system1}-\ref{autonomous-system6}) with the initial conditions $x = 10^{-5}$, $y = 9 \times 10^{-13}$, $u = 10^{-5}$, $\rho = \sqrt{0.999661}$, and $\lambda = 0.8$.\\

In FIG.~\ref{fig:case3_phaseB5}, we observe six critical points in the phase space for the condition $\beta^2 \gamma^2 > 3$. For $\gamma = 1.25$ and $\beta = 1.5$, the critical points $B_7$ and $B_8$ do not exist and the critical point $B_9$ lies outside the accelerating region (magenta/shaded). The critical points $B_5$, $B_6$, and $B_9$ exhibit saddle behavior (unstable) and correspond to the decelerating phase of the Universe. Notably, the point $B_9$ consistently lies outside the accelerating region (magenta/shaded) and thus cannot represent an inflationary solution. There are two heteroclinic orbits: $B_{3\pm} \to B_1 \to B_5$.
\\

In FIG.~\ref{fig:case3_phaseB7}, we observe seven critical points in the phase space under $\beta^2 \gamma^2 > 4$. For $\gamma = 1.4$ and $\beta = 1.5$, the critical point $B_9$ lies outside the accelerating region (magenta/shaded). At these parameter values, the critical points $B_5$, $B_7$, and $B_9$ exhibit saddle-like (unstable) behavior, reflecting the decelerating phase of the Universe. Importantly, point $B_9$ consistently lies outside the accelerating region (magenta/shaded) and never corresponds to an inflationary solution. Two heteroclinic orbits are present: $B_{3\pm} \to B_1 \to B_5$.
 \\

\begin{figure}
     \centering
     \begin{subfigure}[b]{0.4\textwidth}
         \centering
         \includegraphics[width=\linewidth]{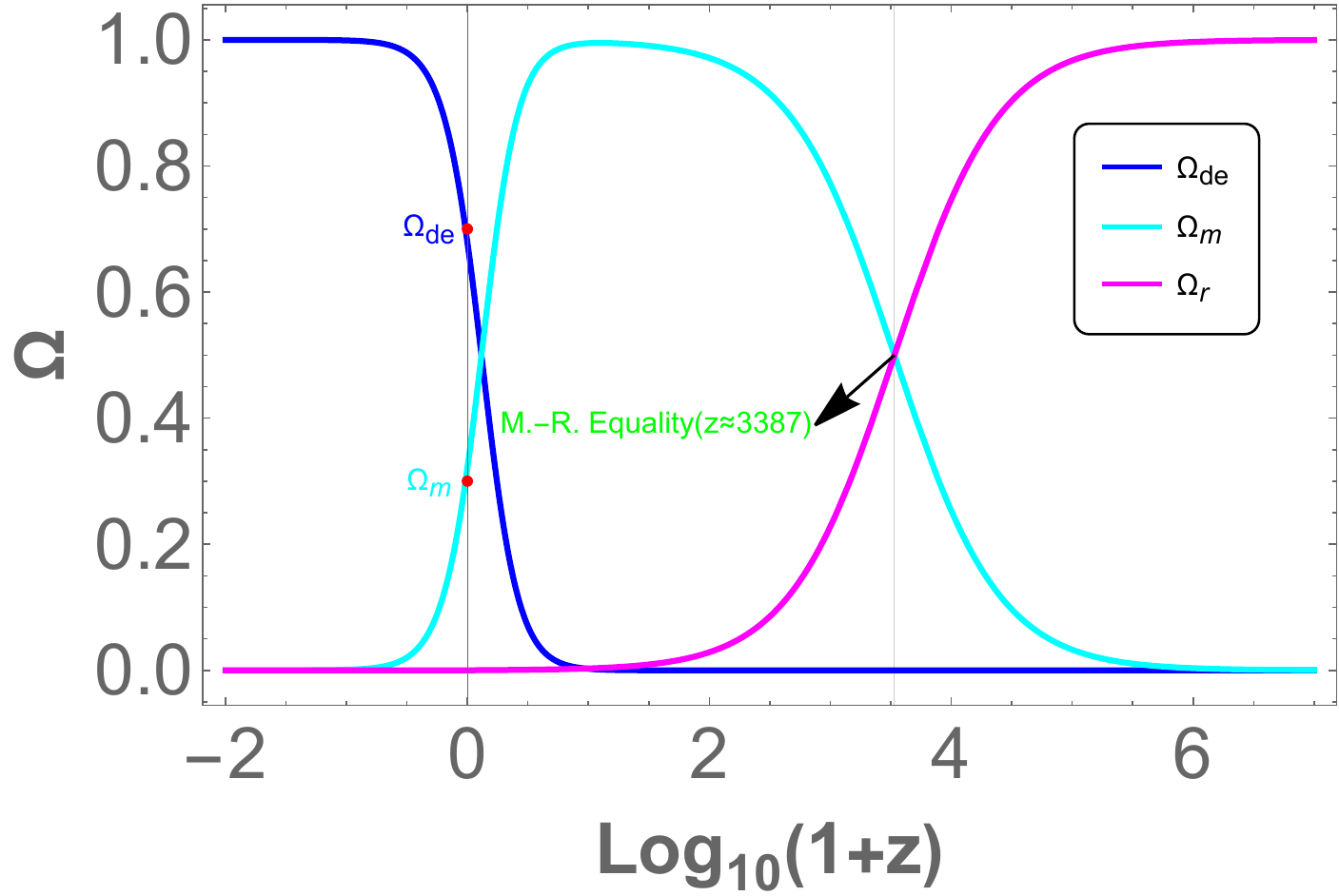}
         \caption{Evolution of density parameters.}
         \label{fig:case3_density}
     \end{subfigure}
     \hfill
     \begin{subfigure}[b]{0.4\textwidth}
         \centering
         \includegraphics[width=\linewidth]{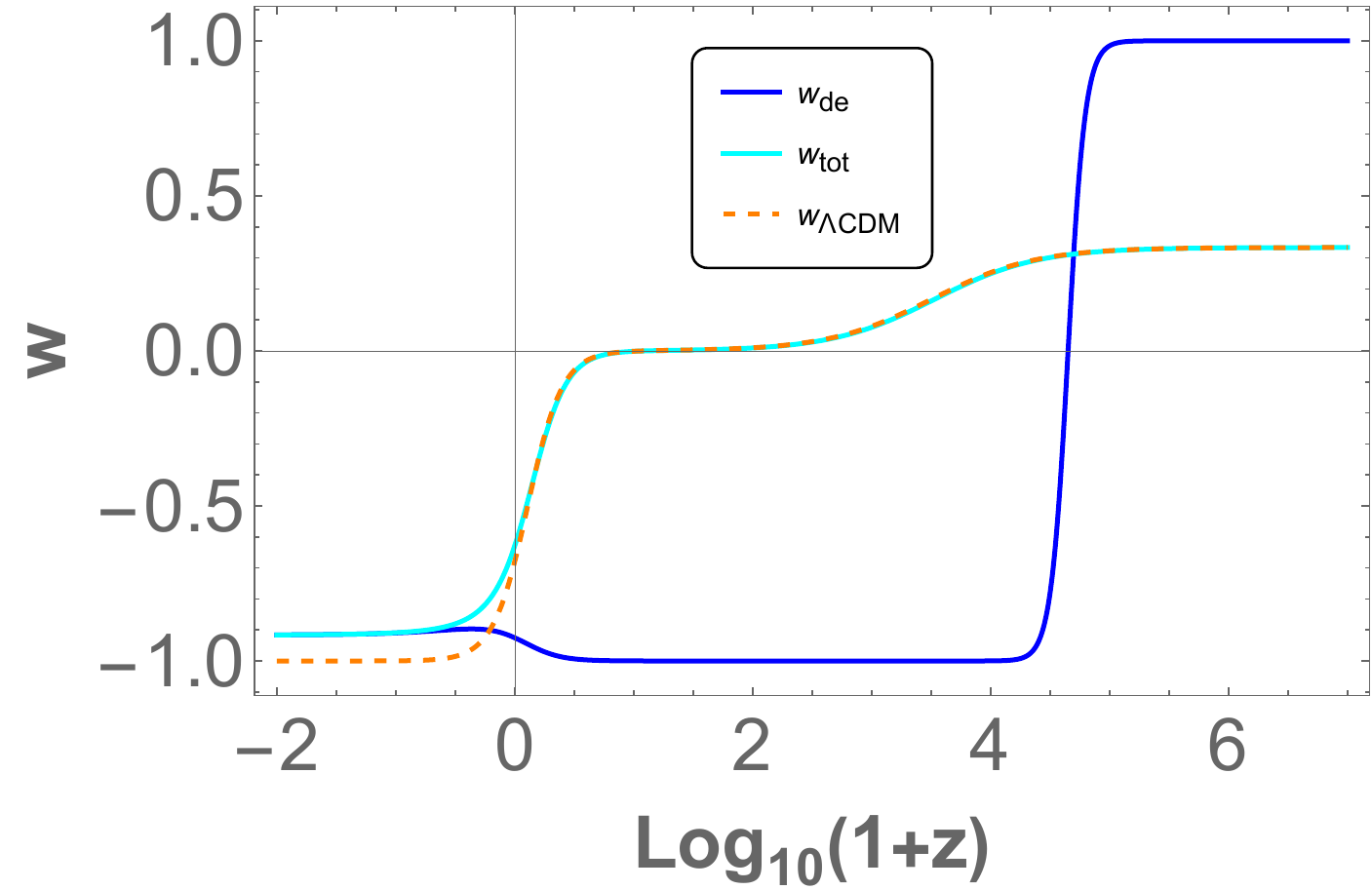}
         \caption{Evolution of EoS parameters in redshift.}
         \label{fig:case3_statepara}
     \end{subfigure}
\caption{In this figure, we set  $\alpha=-0.1$, $\gamma=0.5$ and $\beta=1$ with the initial conditions are the same as in FIG.- \ref{Fig6phase_portrait}.} 
\label{Fig5}
\end{figure}

\begin{figure}
     \centering
     \begin{subfigure}[b]{0.4\textwidth}
         \centering
         \includegraphics[width=\linewidth]{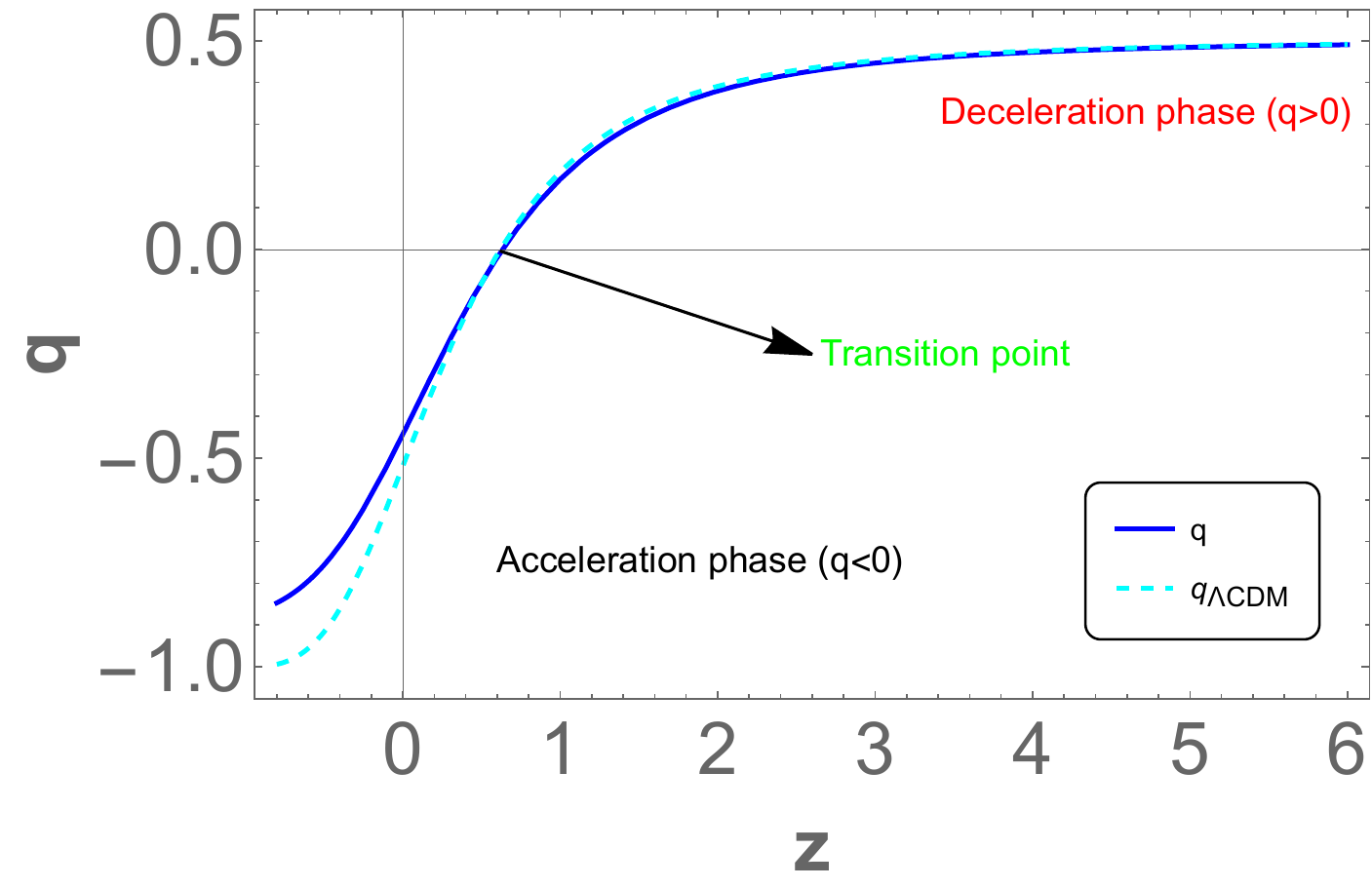}
         \caption{Evolution of the deceleration parameter $q$ in redshift.}
         \label{fig:case3_qz}
     \end{subfigure}
     \hfill
     \begin{subfigure}[b]{0.4\textwidth}
         \centering
         \includegraphics[width=\linewidth]{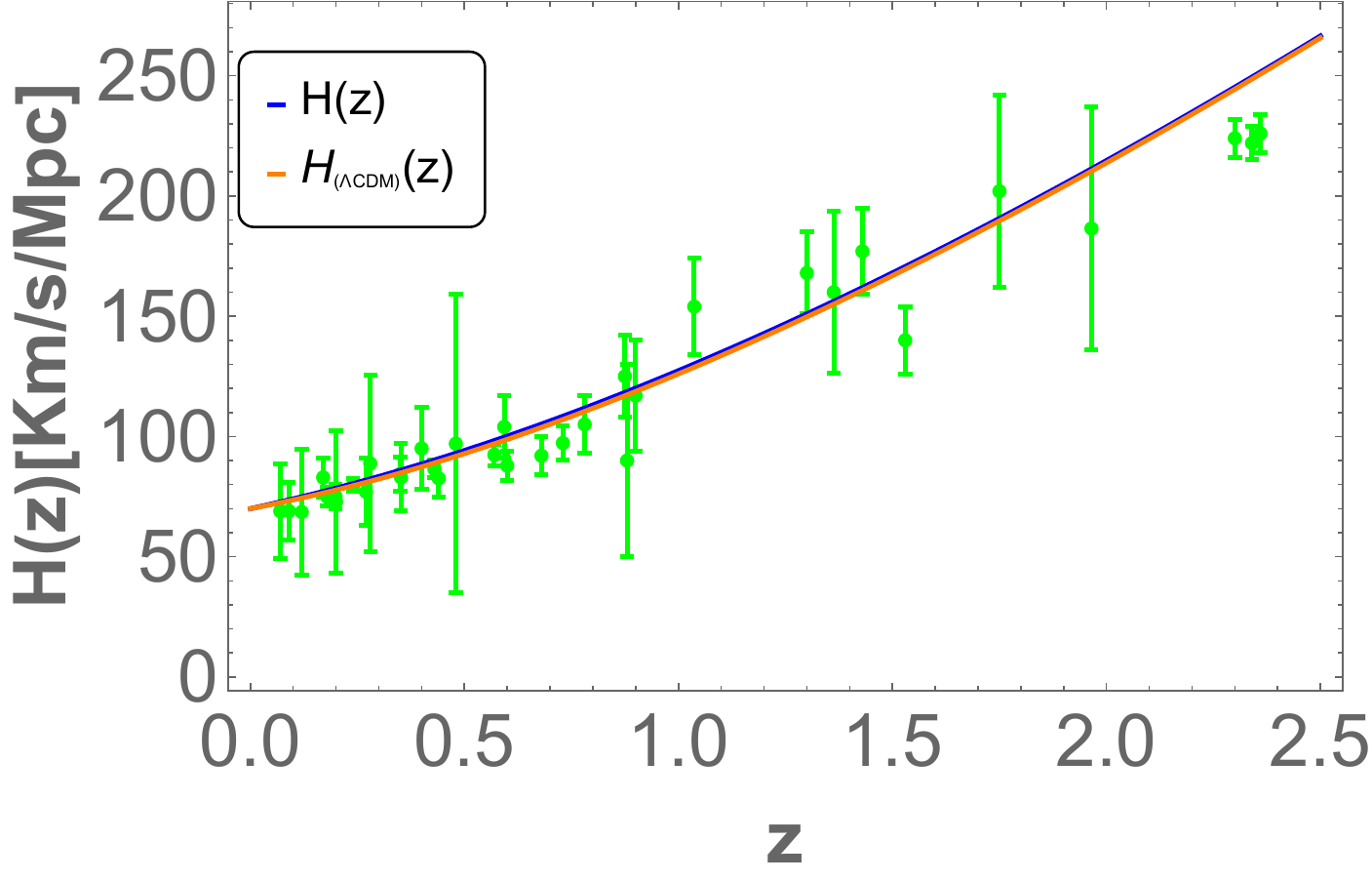}
         \caption{Evolution of the Hubble rate $H(z)$ in redshift.}
         \label{fig:case3_Hz}
     \end{subfigure}
\caption{In this figure, we set  $\alpha=-0.1$, $\gamma=0.5$ and $\beta=1$ with the initial conditions are the same as in FIG.- \ref{Fig6phase_portrait}.} 
\label{Fig6}
\end{figure}

\begin{figure}[H]
 \centering
 \includegraphics[width=90mm]{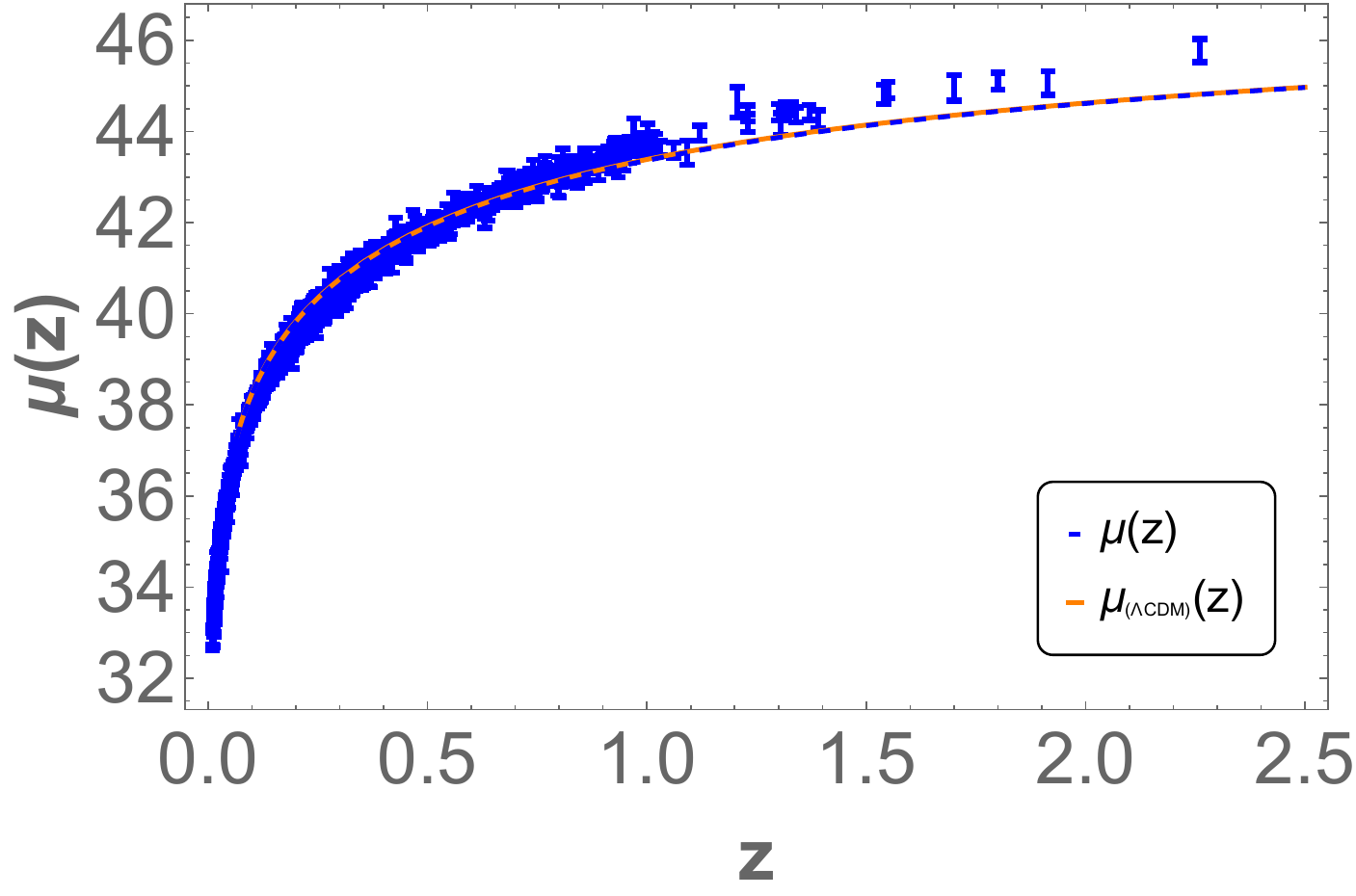} 
 \caption{We show the evolution of the distance modulus function $\mu(z)$ ( dashed blue) and the $\Lambda$CDM model distance modulus function $\mu_{\Lambda CDM}(z)$ along with the 1048 Supernovae Ia (SNe Ia)  data points \cite{Scolnic_2018}. In this figure, we set  $\alpha=-0.1$, $\gamma=0.5$ and $\beta=1$ with the initial conditions are the same as in FIG.- \ref{Fig6phase_portrait}.} \label{Fig6pan}
 \end{figure}

 For these solutions to be cosmologically viable, it is essential to have a sufficiently flat potential ($\beta^2 \gamma^2 < 2$) and to finely tune the initial conditions to ensure the persistence of dark energy domination. The solution should follow the sequence: $B_{3\pm} \to B_1 \to B_9$ ( FIG.~\ref{fig:case3_phaseportrait}). Moreover, at early times, the only viable solutions correspond to non-physical stiff fluid Universe, which are inadequate for describing the Universe. The EoS parameters, energy densities, deceleration parameters, Hubble rate and modulus function for a solution that mirrors the heteroclinic sequence $B_{3\pm} \to B_1 \to B_9$ are illustrated in FIG.-\ref{Fig5}, FIG.-\ref{Fig6} and FIG.- \ref{Fig6pan}.
\\

From FIG.- \ref{fig:case3_density}, one can observe the radiation at the early epoch followed by dominance of dark matter for a while and finally the dark energy phase. At present, $\Omega_{m}\approx 0.32$ and $\Omega_{de}\approx 0.68$, the matter-radiation equality observed at $z\approx 3387$. FIG.-\ref{fig:case3_statepara} explains the EoS parameter $\omega_{tot}$ (cyan), which begins at $\frac{1}{3}$ for radiation falls to $0$ during the matter-dominated period and finally reaches approximately $-1$.  We observed the current value of $\omega_{de}(z=0)=-0.93$. The deceleration parameters [FIG.-\ref{fig:case3_qz}] show the transition from deceleration to acceleration at $z\approx 0.60$. At present, the deceleration parameter becomes $\approx -0.44$. The Hubble rate evolution as a function of redshift $z$, the Hubble rate $H_{\Lambda CDM}(z)$, and the Hubble data points \cite{Moresco_2022_25} are displayed in FIG.-\ref{fig:case3_Hz}. It has been observed that the model presented here closely resembles the standard $\Lambda$CDM model. The $\Lambda$CDM model modulus function $\mu_{\Lambda CDM}$, 1048 pantheon data points, and the evolution of the modulus function $\mu(z)$ are shown in FIG.- \ref{Fig6pan}.\\

\section{Discussions and Conclusion} \label{SECIV}

The dynamical systems technique is an important approach to investigating background cosmology. This provides a way to explore the critical points connected to a model and the characteristics of each point. These points may be associated with observable cosmology and are supported by the evolutionary behavior of the Universe. It may be one of the powerful first tests for any model derived from modified gravity. Further evidence to support or disprove certain models or parameter ranges within the chosen models may be obtained by linking the critical point analysis with their stability and the phase space pictures. Hence, in this paper we have presented the dynamic behavior of the modified Galileon cosmology which is based on higher derivatives in action and requires maintaining second-order equations of motion and imposition of Galilean symmetry. On the other hand,  the constant coefficients of the various action terms in the basic Galileon formulation are further expanded into any functions of the scalar field. We performed the phase space analysis for the exponential form of $F(\phi)$ with three well-motivated functions of potential $V(\phi)$. The cosmological parameters, such as EoS parameters, density parameters and deceleration parameters are obtained through the dynamical variables to analyze the cosmological features of the models.

In all three cases [\ref{case1exp}, \ref{case2power}, \ref{case3sinh}], we have obtained the stable critical points which can describe the late-time cosmic accelerated phase of the Universe. We have also obtained scaling solutions for critical points. The points show that non-standard matter and radiation dominated eras of the Universe. We observe that the results obtained are similar to the standard quintessence model. We can conclude that the Galileon term has no contribution to the dark energy sector of the Universe. But there is the contribution of the Galileon term to the inflationary behavior of the Universe \cite{Burrage_2011, Renaux_Petel_2013, Gonzalez_Espinoza_2019, Choudhury_2024}. For Case-I, we have obtained the value of the density parameters of the matter and dark energy sectors at present ($z=0$), which is $\Omega_{m}^{0}\approx 0.3$, $\Omega_{de}^{0}\approx 0.7$ and also obtained the value of the matter-radiation equality at $z\approx 3387$. Similarly, for Case-II and Case-III, we have $\Omega_{m}^{0}\approx 0.33$, $\Omega_{m}^{0}\approx 0.32$, $\Omega_{de}^{0}\approx 0.67$ and $\Omega_{de}^{0}\approx 0.68$ respectively.  At present, we have obtained the value of the dark energy dominated EoS parameter for Case I, Case II, and Case III as $-1$, $-0.92$, and $-0.93$, respectively. The values are compatible with the current observation \cite{Aghanim:2018eyx}.\\

For Case-\ref{case1exp}, through the behavior of the  FIG.-\ref{fig:case1_qz}, we can say that the deceleration parameter shows the transition from deceleration to acceleration phase at $z=0.66$, and the present value of the deceleration parameter is $q(z=0)=-0.55$. For Case-(\ref{case2power}), we got the transition point at $z=0.62$, and the present value of the deceleration parameter is $q(z=0)=-0.45$ FIG.-\ref{fig:case2_qz}. For Case-(\ref{case3sinh}), we obtain the transition point at $z=0.60$, and the present value of the deceleration parameter is $q(z=0)=-0.44$ FIG.\ref{fig:case3_qz}. In all three cases, transition point and current deceleration parameter values matched the cosmological observations \cite{PhysRevD.90.044016a, PhysRevResearch.2.013028}. In all three cases [\ref{case1exp}, \ref{case2power}, \ref{case3sinh}], we also compared our results with the Hubble 31 data points \cite{Moresco_2022_25} and the Supernovae Ia data 1048 data points \cite{Scolnic_2018}. Through the behavior of the FIGs.-[\ref{fig:case1_Hz}, \ref{fig:case2_Hz}, \ref{fig:case3_Hz}], we can say that the results of our model are very close to the standard $\Lambda$CDM model. In FIGs.-[\ref{Fig2pan}, \ref{Fig4pan}, \ref{Fig6pan}], we plotted the modulus function of our models with the standard $\Lambda$CDM model modulus function along with the 1048 Supernovae Ia data points. The results closely resemble the standard $\Lambda$CDM model.\\

The results of this analysis do not favor any particular form of potential. It shows that class potentials can describe the accelerated expansion of the Universe. As a consequence, the choice of potentials remains arbitrarily made. Even though this analysis is done by choosing three different potentials, it can be expanded to include more potentials. We have restricted ourselves to these three potential forms to keep the analysis from becoming unnecessarily lengthy. Furthermore, this general parameterization of $f(\lambda)=\lambda^2 (\Gamma-1)$ does not depend on the particular scalar field dark energy model but on the definitions of $\Gamma$ and $\lambda$. Furthermore, this general parametrization of $f$ can also be used with other scalar field dark energy models, like a quintessence, phantom, coupled dark energy, modified gravity theories, k-essence, and tracker solutions, etc. In all three cases,  the scaling solution is an attractor during the radiation and matter eras but it does not exit the epoch of cosmic acceleration.  \\

This research demonstrates the potential of modified Galileon theory, which should be investigated in greater detail using observational constraint analysis or perturbation theory in the cosmological context. This may provide additional insights into this theory, including their relationships to the cosmic microwave background radiation power spectrum and the large-scale structure of the Universe. 

 
\section*{Appendix}
\subsection{Center Manifold Theory (CMT) for the non-hyperbolic critical point $B_{2}$}\label{Sec-app}

Perko \cite{Perko2001} explains the mathematical framework of CMT. If the eigenvalues of critical points are zero, linear stability theory cannot explain their stability. In a center manifold theory, the dimensionality of a system near that point is reduced, allowing stability to be examined. When the system passes through the fixed point, it behaves in an invariant local center manifold $W^{c}$. System stability is determined by how stable the reduced system is at a given point.\\
Assume $f \in C^{r}(E)$, where $E$ is an open subset of $R^{n}$ containing the origin and $r  \geq 1$. Suppose $f(0) = 0$ and $Df(0)$ have c eigenvalues with zero real parts and s eigenvalues with negative real parts, where c + s = n. In general, the system is reduced to the following form

\begin{eqnarray}
\dot{x}= Ax + F(x,y) \nonumber \\ 
\dot{y}=  By + G(x,y) \label{CMT1}    
\end{eqnarray}

where $(x, y) \in R^{c} \times R^{s} $, $A$ is a square matrix with c eigenvalues having zero real parts, B is a square matrix with s eigenvalues with negative real parts, and $F(0) = G(0) = 0, DF(0) = DG(0) = 0$. Furthermore, there exists a $\delta > 0$ and a function $h(x) \in C^{r}(N_{\delta}(0))$ that defines the local center manifold and satisfies 

\begin{eqnarray}
\mathcal{N}(h(x))=Dh(x)[Ax+F(x,h(x))] 
 -Bh(x)-G(x,h(x))=0 \label{CMT2}    
\end{eqnarray}

for $|x|< \delta$; and the center manifold  is defined by the
system of differential equations

\begin{eqnarray}
 \dot{x}= Ax + F(x,h(x)) \label{CMT3}   
\end{eqnarray}

for all $x\in R^{c}$ with $|x|<\delta$.\\

The Jacobian matrix at the critical point $B_{2}$ for the autonomous system (\ref{autonomous-system1}-\ref{autonomous-system6}) is as follows:

\[
J(B_{2}) = 
\left( \begin{array}{ccccc}
 -3 & 0  & 0  & 0 & \sqrt{\frac{3}{2}} \\
  0 & -3 & 0 & 0 & 0 \\
  0 & 0 & -3 & 0 & 0 \\
  0 & 0 & 0 & -2 & 0 \\
  0 & 0 & 0 & 0 & 0 
\end{array} \right)
\]

The eigenvalues of Jacobian matrix $J(B_{2})$ are $-3$, $-3$, $-3$, $-2$, and $0$. The \begin{math}\left[1, 0, 0, 0, 0 \right]^T \end{math}, \begin{math}\left[0, 1, 0, 0, 0 \right]^T \end{math}, \begin{math}\left[0, 0, 1, 0, 0 \right]^T \end{math} are the eigenvector to the corresponding eigenvalues $-3$, and \begin{math} \left[0,0,0,1,0\right]^{T}\end{math} be the eigenvector corresponding to the eigenvalue $-2$ and \begin{math} \left[0,0,0,0,0\right]^{T}\end{math} be the eigenvector corresponding the eigenvalue $0$.\\

Using center manifold theory, we examine the stability of the critical point $B_{2}$ because of its non-hyperbolic nature. To apply CMT to this critical point, we must shift it to the origin using a shifting transformation. we have followed these transformations: $X=x$, $Y=1-y$, $Z=u$, $R=\rho$ and $ S=\lambda$, and then we can write equations in the new coordinate system as 

\begin{eqnarray}
\left( \begin{array}{c}
\dot{X} \\ 
\dot{Y} \\ 
\dot{Z} \\ 
\dot{R} \\ 
\dot{S} 
\end{array} \right) &=& 
\left( \begin{array}{ccccc}
-3 & 0 & 0 & 0 & 0 \\
0 & -3 & 0 & 0 & 0 \\
0 & 0 & -3 & 0 & 0 \\
0 & 0 & 0 & -2 & 0 \\
0 & 0 & 0 & 0 & 0
\end{array} \right) 
\left( \begin{array}{c}
X \\ 
Y \\ 
Z \\ 
R \\ 
S    
\end{array} \right) +
\left( \begin{array}{c}
non \\ 
linear \\ 
term   
\end{array} \right)
\end{eqnarray}

Comparing this diagonal matrix with the general form (\ref{CMT1}). After that, we can say that here $X$, $Y$, $Z$, and $R$ are the stable variables, and $S$ is the central variable. At this critical point, the $A$ and $B$ matrix appears as 

\[
A =
\left( \begin{array}{cccc}
 -3 & 0 & 0 & 0 \\
  0 & -3 & 0 & 0 \\
  0 & 0 & -3 & 0 \\
  0 & 0 & 0 & -2
\end{array} \right)
\hspace{0.5cm}
B = 
\left( \begin{array}{c}
 0    
\end{array} \right)
\]

According to CMT, the manifold can be defined by a continuous differential function. we have assumed the following functions for the stable variables $X=h_{1}(S), Y=h_{2}(S), Z=h_{3}(S),$ and  $R=h_{4}(S)$. With the help of the equation (\ref{CMT2}), we have obtained the following zeroth approximation of the manifold functions 

\begin{eqnarray}
&\mathcal{N}(h_1(S)) = -\sqrt{\frac{3}{2}}S\,, \hspace{1cm} \mathcal{N}(h_2(S)) =0\,, \nonumber \\&  \mathcal{N}(h_3(S)) =0 \,, \hspace{1cm} \mathcal{N}(h_4(S)) =0.
\end{eqnarray}

As of this moment, the center manifold acquired by 
\begin{equation}\label{CMT-manifold}
\dot{S}= 3 S^{3}(\Gamma-1)+ higher \hspace{0.2cm} order \hspace{0.2cm} term   
\end{equation}

According to the CMT, this critical point shows stable behavior for  $\Gamma<1$ and unstable for $\Gamma>1$, where $\Gamma$ depends on the potential function $V(\phi)$.
\subsection{Datasets} \label{dataset}
\subsubsection*{\bf{Hubble data $H(z)$}}
This study will examine 31 data points \cite{Moresco_2022_25} to study the behavior of the Hubble rate of our model. We have also compared our model with the standard $\Lambda$CDM model. The standard $\Lambda$CDM can be define as 
\begin{equation}\label{hubble_LCDM}
H_{\Lambda CDM}= H_{0}\sqrt{(1+z)^3 \Omega_{m}+(1+z)^4 \Omega_{r}+\Omega_{de}} \,,   
\end{equation}

\subsubsection*{\bf{Supernovae Ia}}
A comprehensive dataset for analyzing Type Ia supernovae, the Pantheon collection combines information from many observatories and surveys. There are 1048 data points in this collection, which range in redshift from 0.01 to 2.3 \cite{Scolnic_2018}. It includes observations from well-known programs such as the Panoramic Survey Telescope and Rapid Response System (PanSTARRS1), Hubble Space Telescope (HST) survey, Super-Nova Legacy Survey (SNLS), and Sloan Digital Sky Survey (SDSS). The Pantheon collection, which brings together data from several sources, provides a significant understanding of the characteristics and actions of Type Ia supernovae and their cosmic relevance. The luminosity of a star is an alternative method of defining distances in an expanding Universe. The distance moduli function can be defined as 
\begin{equation}\label{panmoduli}
\mu(z_{i}, \Theta)=5 \log_{10}[D_{L}(z_i, \Theta)]+M    
\end{equation}
where M and $D_{L}$  represented the nuisance parameter and the luminosity distance, respectively. The luminosity distance can be written as 
\begin{equation}\label{luminositydistance}
D(z_{i}, \Theta)=c (1+z_{i}) \int^{z_i}_{0} \frac{dz}{H(z, \Theta)} \,,   
\end{equation}

\section*{Acknowledgements} LKD acknowledges the financial support provided by the University Grants Commission (UGC) through Senior Research Fellowship UGC Ref. No.: 191620180688. BM acknowledges Bauman Moscow State Technical University (BMSTU), Moscow, for providing the Visiting Professor position during which part of this work has been accomplished. The authors are thankful to the honorable reviewers for their constructive comments and suggestions to improve the quality of the paper.

\section*{References}
\bibliographystyle{utphys}
\bibliography{ref_short}

\end{document}